\documentclass[11pt]{article}

\usepackage{amssymb}
\usepackage{amsmath}
\usepackage{fullpage}
\usepackage{times}
\usepackage{graphicx}
\newcommand{\SFW}{\mbox{S$^4$W}}
\newcommand{\paper}{paper}

\title{Using Hierarchical Data Mining to Characterize \\
Performance of Wireless System Configurations}
\author{Alex Verstak$^*$, Naren Ramakrishnan$^*$, Kyung Kyoon Bae$^{\dagger}$, William H. Tranter$^{\dagger}$,\\
Layne T. Watson$^*$, Jian He$^*$, and Clifford A. Shaffer$^*$\\
\large $^*$Department of Computer Science\\
\large $^{\dagger}$Bradley Department of Electrical and Computer Engineering\\
\large Virginia Polytechnic Institute and State University\\
\large Blacksburg, VA 24061\\
\,\,\,\,\\
\large Theodore S. Rappaport\\
\large Department of Electrical and Computer Engineering\\
\large University of Texas\\
\large Austin, TX 78712}

\date{}
\begin{document}

\maketitle

\begin{abstract}
\noindent
This \paper{} presents a statistical framework for assessing wireless
systems performance using hierarchical data mining techniques. We
consider WCDMA (wideband code division multiple access) systems with
two-branch STTD (space time transmit diversity) and 1/2 rate
convolutional coding (forward error correction codes). Monte Carlo
simulation estimates the bit error probability (BEP) of the system
across a wide range of signal-to-noise ratios (SNRs).  A performance
database of simulation runs is collected over a targeted space of
system configurations.  This database is then mined to obtain regions
of the configuration space that exhibit acceptable average performance.
The shape of the mined regions illustrates the joint influence of
configuration parameters on system performance.  The role of data
mining in this application is to provide explainable and statistically
valid design conclusions.  The research issue is to define
statistically meaningful aggregation of data in a manner that permits
efficient and effective data mining algorithms. We achieve a good
compromise between these goals and help establish the applicability of
data mining for characterizing wireless systems performance.
\end{abstract}
\thispagestyle{empty}
\newpage

\section{Introduction}

Data mining is becoming increasingly relevant in simulation methodology
and computational science~\cite{naren-ayg}.  It entails the
`non-trivial process of identifying valid, novel, potentially useful,
and ultimately understandable patterns in data'~\cite{kdd-cacm}.  Data
mining can
be used in both predictive (e.g., quantitative assessment of factors on
some performance metric) and descriptive (e.g., summarization and
system characterization) settings. Our goal in this \paper{} is to
demonstrate a hierarchical data mining framework applied to the problem
of characterizing wireless system performance.

This work is done in the context of the \SFW{} problem solving
environment~\cite{ipdps-s4w}---`Site-Specific System Simulator for
Wireless System Design'.  \SFW{} provides site-specific (deterministic)
electromagnetic propagation models as well as stochastic wireless
system models for predicting the performance of wireless systems in
specific environments, such as office buildings. \SFW{} is also
designed to support the inclusion of new models into the system,
visualization of results produced by the models, integration of
optimization loops around the models, validation of models by
comparison with field measurements, and management of the results
produced by a large series of experiments.  In this paper,
we study the effect of
configuration parameters on the bit error probability (BEP) of a system
simulated in~\SFW.

The approach we take is to accumulate a performance database of
simulation runs that sweep over a targeted space of system
configurations. This database is then mined to obtain regions of the
configuration space that exhibit acceptable average performance.
Exploiting prior knowledge about the underlying simulation, organizing
the computational steps in data mining, and interpreting the results at
every stage, are important research issues. In addition, we bring out
the often prevailing tension between making statistically meaningful
conclusions and the assumptions required for efficient and effective
data mining algorithms. This interplay leads to a novel set of problems
that we address in the context of the wireless systems performance
domain.

Data mining algorithms work in a variety of ways but, for the purposes
of this \paper{}, it is helpful to think of them as performing systematic
aggregation and redescription of data into higher-level objects.  Our
work can be viewed as employing three such layers of aggregation:
points, buckets, and regions. Points (configurations) are records in
the performance database. These records contain configuration
parameters as well as unbiased estimates of bit error probabilities
that we use as performance metrics.  Buckets represent averages of
points.  We use buckets to reduce data dimensionality to two, which is
the most convenient number of dimensions for visualization.  Finally,
buckets are aggregated into 2D regions of constrained shape.  We find
regions of buckets where we are most confident that the configurations
exhibit acceptable average performance.  The shapes of these regions
illustrate the nature of the joint influence of the two selected
configuration parameters on the configuration performance.  Specific
region attributes, such as region width, provide estimates for the
thresholds of sensitivity of configurations to variations in parameter
values.

\subsection{Reader's Guide}

Our major contribution is the development of a statistical framework
for assessing wireless system performance using data mining
techniques.  The following section outlines wireless systems
performance simulation methodology and develops a statistical framework
for spatial aggregation of simulation results.
Section~\ref{sec:w-example} demonstrates a substantial subset of this
framework in the context of a performance study of WCDMA (wideband code
division multiple access~\cite{wcdma-holma}) systems that employ
two-branch STTD (space-time transmit diversity~\cite{sttd-alamouti})
techniques and 1/2 rate convolutional coding (forward error correction
codes~\cite{wcdma-holma}).  We study the effect of power imbalance
between the branches on the BEP of the system across a wide range of
average signal-to-noise ratios (SNRs).  Section~\ref{sec:gizmo} extends
the statistical framework to support computation of optimized regions
of the bucket space.  Such regions are computed by a well-known data
mining algorithm~\cite{fukuda-tods, fukuda-rectilinear}.
Section~\ref{sec:experiments} applies these concepts to the example in
Section~\ref{sec:w-example}.  Section~\ref{sec:conclusion} summarizes
present findings and outlines directions for future research.

\section{The Statistics of Aggregation and the Aggregation of Statistics}
\label{sec:stat}

Temporal variations in wireless channels have been extensively studied
in the literature~\cite{comm-ziemer}.  The present work uses a Monte
Carlo simulation of WCDMA wireless systems to study the effect of these
variations.  The simulation traces a number of frames of random
information bits through the encoding filters, the channel (a Rayleigh
fading linear filter~\cite{cm-hashemi}), and the decoding filters.  The
inputs are hardware parameters, average SNR, channel impulse response,
and the number of frames to simulate.  The output is the bit error
rate---the ratio of the number of information bits decoded in error to
the total number of information bits simulated.  Simulations of this
kind statistically model channel variations due to changes in the
environment and device movement across a small geographical area
(\emph{small-scale fading}~\cite{cm-hashemi}).  We refer to this kind
of channel variation as \emph{temporal variation} because a system is
simulated over a period of time.  Further, we say that a given list of
inputs to the WCDMA simulation is a \emph{configuration} or a
\emph{point} in the configuration space.

\begin{figure}
\begin{center}
\includegraphics[width=4.0in]{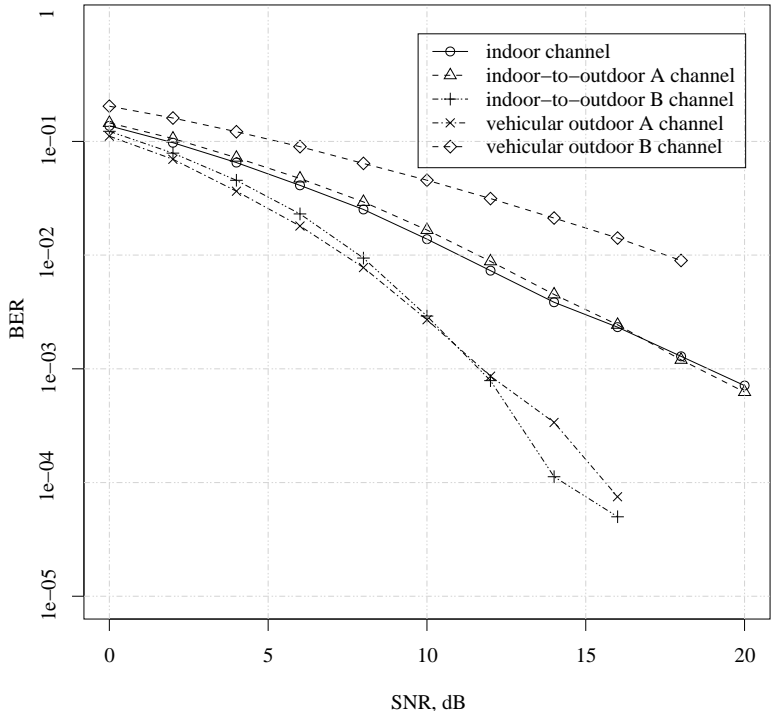}
\end{center}
\caption[Typical 1D slices of the configuration space.]{Typical 1D
slices of the configuration space.  The plots show simulated BERs (bit
error rates) of wireless systems for five common benchmark
channels~\cite{umts-utra} across a typical range of average SNRs.}
\label{fig:2d-example}
\end{figure}

\emph{Spatial variations} are due to changes in system configurations.
We use this term to describe two quite different phenomena: changes in
the average SNR and channel impulse response due to \emph{large-scale
fading}~\cite{cm-hashemi} and variations of hardware parameters.  A
typical approach to the analysis of spatial variations is to run
several temporal variation simulations (i.e., compute bit error
rates --- BERs --- at several
points within a given area of interest) and plot 1D or 2D slices of the
configuration space, as shown in Figure~\ref{fig:2d-example}.  In this
\paper{}, we augment this approach with statistically meaningful
aggregation of performance estimates across several points.  The result
of this aggregation is a space of buckets, each bucket representing the
aggregation of a number of points. Moving up one level of aggregation
in this manner allows us to bring data mining algorithms to operate at
the level of buckets. The space of buckets mined by the data mining
algorithm is then visualized using color maps. The color of each bucket
is the confidence that the points (configurations) that map to this
bucket exhibit acceptable average performance.

\begin{table}
\begin{center}
\begin{tabular}{c l}
\hline
$c,C,R$ & entities (points, buckets, regions) \\
$x,b,B$ & random variables \\
$E[x],E[b],E[B]$ & true means of random variables $x$, $b$, $B$ \\
$\sigma^2,\Sigma^2$ & true variances of random variables $b$, $B$ \\
$\hat{x},\hat{b},\hat{B}$ & estimates of means $E[x]$, $E[b]$, $E[B]$ of random variables $x$, $b$, $B$ \\
$\hat{\sigma}^2,\hat{\Sigma}^2$ & estimates of variances $\sigma^2$, $\Sigma^2$ of random variables $b$, $B$ \\
$P(E)$ & probability of event $E$, where $E$ is a boolean condition \\
$F_{N-1}(T)$ & $P(X<T)$ for $X$ having the Student~$t$ distribution with $N-1$ degrees of freedom \\
$\{x_k\}_{k=1}^n$ & set $\{x_1,x_2,\ldots,x_n\}$ \\
$\{\{x_{kj}\}_{j=1}^{n_k}\}_{k=1}^n$ & set $\{x_{11},x_{12},\ldots,x_{1n_1},x_{21},x_{22},\ldots,x_{2n_2},\ldots,x_{n1},x_{n2},\ldots,x_{nn_n}\}$ \\
\hline
\end{tabular}
\end{center}
\caption[Summary of mathematical notation.]{Summary of mathematical
notation. Lower case letters are used for points and upper case letters
are used for buckets and regions.  Additional conventions are
introduced in Table~\ref{tab:notation2}.}
\label{tab:notation}
\end{table}

\subsection{The First Level of Aggregation: Points}

Table~\ref{tab:notation} summarizes some of the syntactic conventions
used in this \paper{}.  Mathematically, we can think of the WCDMA
simulation as estimating the mean~$E[x_k]$ of a random variable~$x_k$
with some (unknown) distribution~\cite{comm-jeruchim} ($x_k$ is one
when the information bit is decoded in error or zero when it is decoded
correctly).  Each BER $\hat{x}_{kj}$, $1\le{}j\le{}n_k$, output by the
simulation is an unbiased estimate of the BEP~$E[x_k]$ of the simulated
configuration~$c_k$.  
Instead of building a detailed stochastic model of the
simulation (analytically, from the
distribution of $x_k$), we choose to work with the simpler distribution of 
the BER $\hat{x}_{kj}$, referred to henceforth as just $b_k$.
Thus, each sample from the distribution of $b_k$ is realized by
simulating a number of frames and obtaining an estimate
of $E[x_k]$.
The distribution of~$b_k$ is
approximately Gaussian due to the Central Limit Theorem.  
Technically, we assume
that the number of frames per estimate $\hat{x}_{kj}$ is `large enough'
so that the Lindeberg condition is satisfied, that the variance
of~$\hat{x}_{kj}$ is finite, and that $\{\hat{x}_{kj}\}_{j=1}^{n_k}$
are i.i.d.  We say that $E[b_k]=E[E[x_k]]$ is the \emph{expected BEP}
of configuration~$c_k$ under Rayleigh fading.

\subsection{The Second Level of Aggregation: Buckets}

Let us now aggregate several points (i.e., random variables) into one
bucket.  The purpose of this aggregation is to reduce data
dimensionality to a size that is easy to visualize, usually one or two
dimensions.  The basic idea is to linearly average all points that map
to the same bucket but we must do so carefully, in order to preserve a
meaningful statistical interpretation.  Let $\{b_k\}_{k=1}^n$ be
Gaussian random variables with means $\{E[b_k]\}_{k=1}^n$ and variances
$\{\sigma^2_k\}_{k=1}^n$.  As in the previous paragraph, let each such
variable~$b_k$ be the estimated BEP of some configuration~$c_k$,
$1\le{}k\le{}n$.  For bucket~$C$, define a \emph{bucket} (mixture)
\emph{random variable}~$B$ as the convex combination 
$$B=\sum_{k=1}^np_kb_k,$$ where
the $p_k \geq 0$ and $\sum_{k=1}^{n} p_k = 1$.
It is convenient to make $\{p_k\}_{k=1}^n$ the
probabilities of occurrence of the configurations $\{c_k\}_{k=1}^n$ in
the dataset being analyzed.  This setup underlines the dependence of
the outputs on the distribution of the inputs and frees the user from
having to provide values for the constants $\{p_k\}_{k=1}^n$.  It is
well known that, as long as $\{b_k\}_{k=1}^n$ are \emph{mutually
independent} and Gaussian with means~$\{E[b_k]\}_{k=1}^n$ and
variances~$\{\sigma^2_k\}_{k=1}^n$, $B$ is Gaussian with mean
$E[B]=\sum_{k=1}^np_kE[b_k]$ and variance
$\Sigma^2=\sum_{k=1}^np_k^2\sigma_k^2$~\cite{stat-casella}.  The
expected value $E[B]$ of the random variable~$B$ can be viewed as the
expected BEP of bucket $C=\{c_1,c_2,\ldots,c_n\}$ in a Rayleigh fading
environment, conditional on the (discrete) distribution of the
configurations in~$C$.

The values $\{p_k\}_{k=1}^n$ are what the statisticians call
\emph{prior probabilities}.  For most purposes of this \paper{}, we simply
estimate $\{p_k\}_{k=1}^n$ from available data.  These values are
explicitly or implicitly constructed during experiment design and we
assume that they remain constant during experiment analysis. However,
one can collect additional data as long as doing so does not change
$\{p_k\}_{k=1}^n$.  Prior probabilities can come from a number of
sources:  channel sounding measurements, propagation simulations,
hardware and budget constraints, or even educated guesses by wireless
system designers.  The rest of the \paper{} silently assumes that the
values $\{p_k\}_{k=1}^n$ have been established beforehand.  It is
important to remember that even though the prior probabilities are for
the most part transparent to the analysis presented here, they
nonetheless always exist and all conclusions of data analysis are made
conditional on the prior probabilities.

This discussion of $\{p_k\}_{k=1}^n$ can be interpreted as a deferral
of the exact definition of~$B$ until experiment setup, or as
parameterization of the analysis procedure.  A natural question is
whether or not this level of parameterization is sufficient.  It is
sufficient for the purposes of this \paper{} but, strictly speaking,
the interrelations between $\{b_k\}_{k=1}^n$ should also be defined
during experiment setup.  Mutual independence of $\{b_k\}_{k=1}^n$ is a
simplifying assumption and it might be desirable to model interactions
between $\{b_k\}_{k=1}^n$ in practice.  This implies adding covariance
terms to~$\Sigma^2$ and re-thinking the distribution of~$B$.  Such
analysis is necessarily specific to a particular experiment.  For the
sake of simplicity, the rest of this \paper{} assumes mutual
independence of variables in a given bucket.

\subsection{Confidence Estimation}

Point and bucket estimates of the expected BEP are meaningful
performance metrics for wireless systems.  Let us also estimate our
confidence in these estimates.  Confidence analysis enables wireless
system designers to make more practical claims than point estimates
alone.  A statement of the form `this configuration will exhibit
acceptable performance in 95\% of the cases' is often preferable to a
statement of the form `the expected BEP of this configuration is
approximately $5\times{}10^{-4}$'.  More precisely, we say that
configuration~$c_k$ \emph{exhibits acceptable performance} when the
expected BEP $E[b_k]$ of configuration $c_k$ is below some fixed
threshold~$T$.  This statement is conditional on the temporal
simulation assumptions, i.e., Rayleigh fading.  Standard values for~$T$
are $10^{-3}$ for voice quality systems and $10^{-6}$ for data quality
systems.  Likewise, we say that bucket~$C$ (a subspace of
configurations) \emph{exhibits acceptable average performance} when the
expected BEP $E[B]$ of bucket~$C$ is below some fixed threshold~$T$.
This statement is conditional on both the temporal simulation
assumptions and the distribution of configurations $\{c_k\}_{k=1}^n$ in
the bucket (the prior probabilities).

The confidence that configuration~$c_k$ (resp. bucket~$C$) exhibits
acceptable (average) performance is $P(E[b_k]<T)$ (resp.
$P(E[B]<T)$).  Since $b_k$ and~$B$ are Gaussian, these probabilities
can be estimated as
$$P(E[b_k]<T)\approx{}F_{n_k-1}\left(\frac{T-\hat{b}_k}{\hat{\sigma}_k/\sqrt{n_k}}\right),\quad
P(E[B]<T)\approx{}F_{N-1}\left(\frac{T-\hat{B}}{\hat{\Sigma}/\sqrt{N}}\right),$$
where $F_{N-1}(\cdot)$ is the CDF of the Student~$t$ distribution with
$N-1$ degrees of freedom and $n_k$ and~$N$ are the sample sizes for
configuration~$c_k$ and bucket~$C$, respectively.  For
configuration~$c_k$,
$$\hat{b}_k=\frac{1}{n_k}\sum_{j=1}^{n_k}\hat{x}_{kj},\quad
\hat{\sigma}^2_k=\frac{1}{(n_k-1)}\sum_{j=1}^{n_k}(\hat{x}_{kj}-\hat{b}_k)^2,$$
where $\hat{b}_k$ and $\hat{\sigma}_k^2$ are the estimates of the
expected BEP and the BEP variance at point~$c_k$, $n_k\ge{}2$ is sample
size, and $\{\hat{x}_{kj}\}_{j=1}^{n_k}$ are sample values.  For
bucket~$C$, we substitute point estimates into
$E[B]=\sum_{k=1}^np_kE[b_k]$ and $\Sigma^2=\sum_{k=1}^np_k^2\sigma_k^2$
to obtain $$\hat{B}=\sum_{k=1}^n\hat{p}_k\hat{b}_k, \quad
\hat{\Sigma}^2=\sum_{k=1}^n\hat{p}_k^2\hat{\sigma}_k^2,$$
where $\hat{B}$ and $\hat{\Sigma}^2$ are
the estimates of the expected BEP and the BEP variance at bucket~$C$,
and
$\{\hat{p}_k\}_{k=1}^n$ are the prior probabilities estimated from the
dataset as $\hat{p}_k=n_k/\sum_{i=1}^nn_i$.  Observe that
$$\hat{B}=\sum_{k=1}^n\hat{p}_k\hat{b}_k=\frac{1}{\sum_{k=1}^nn_k}\sum_{k=1}^n\sum_{j=1}^{n_k}\hat{x}_{kj}$$
is exactly the sample mean of all observations in the bucket, but
$\hat{\Sigma}^2$ is \emph{not} the variance 
of this sample.  This is the case because
$\{\{\hat{x}_{kj}\}_{j=1}^{n_k}\}_{k=1}^n$ are not i.i.d.  samples from
the mixture distribution of~$B$---they are samples from the constituent
distributions of $\{b_k\}_{k=1}^n$.

\section{Extended Example}
\label{sec:w-example}

Let us now apply the techniques developed so far to analyze the
performance of a space of configurations.  The wireless systems under
consideration employ WCDMA technology with two-branch STTD and
$1/2$~rate convolutional coding.  We require that the transmitter has
two antennas (branches) separated by a distance large enough for their
signals to be uncorrelated, but small enough for the mean path losses
and impulse responses of their channels to be approximately equal at
receiver locations of interest.  We assume Rayleigh flat fading
channels, which is reasonable for indoor applications in the ISM and
UNII carrier frequency bands (2.4 and 5.2~GHz, respectively).  The goal
is to study the effect of power imbalance between the branches on the
BEP of the configurations across a wide range of average SNRs.

This section presents a number of plots that summarize simulated BERs.
We also outline the process of statistically significant sampling of
the configuration space.  The next section develops a data mining
methodology that solves a practically important problem: given a
dataset similar to the one presented next, find a region of the
configuration space where we can confidently claim that configurations
will exhibit acceptable (average) performance.

\begin{figure}
\begin{center}
\includegraphics[width=4.0in]{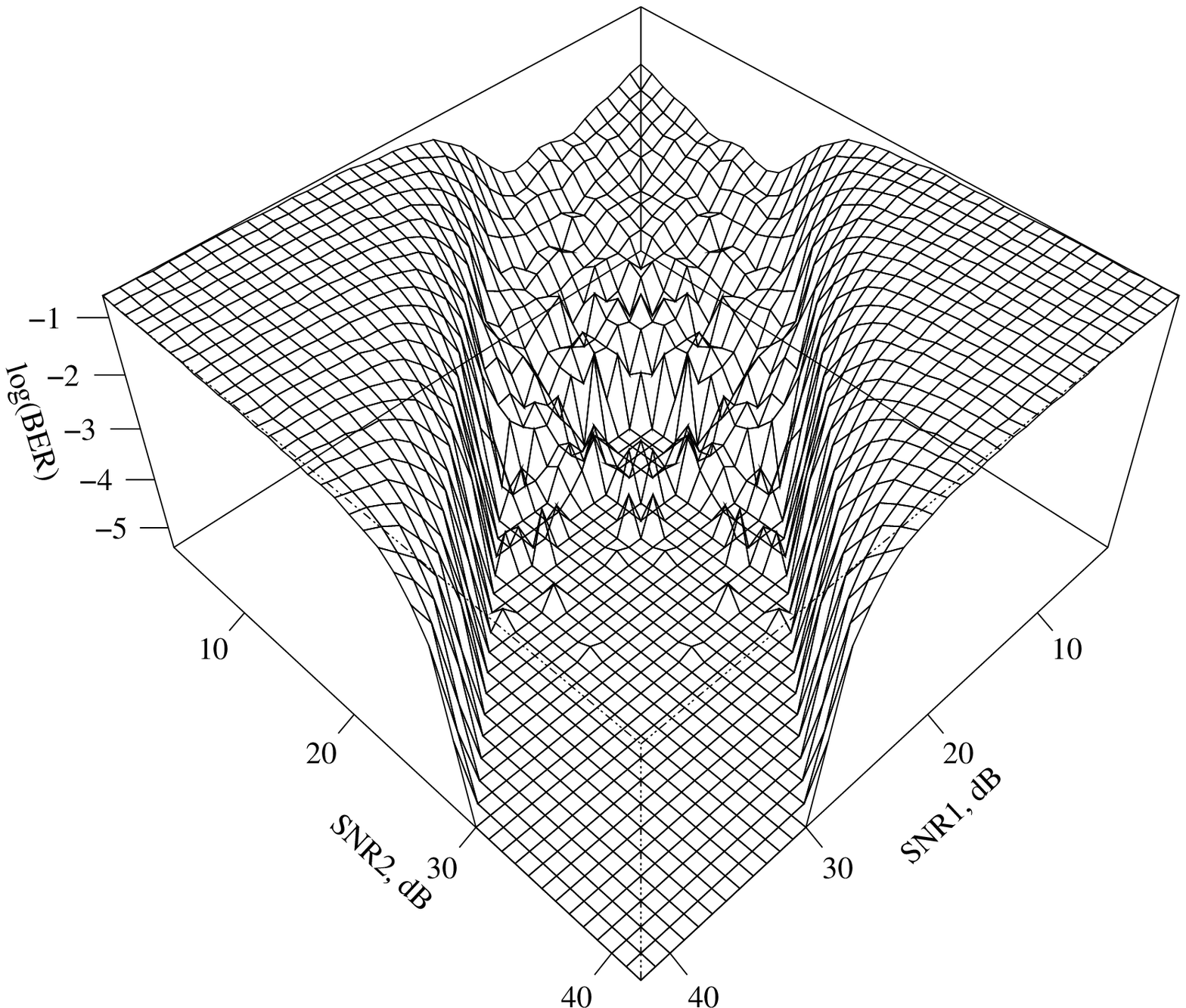} \\
\medskip
\includegraphics[width=4.0in]{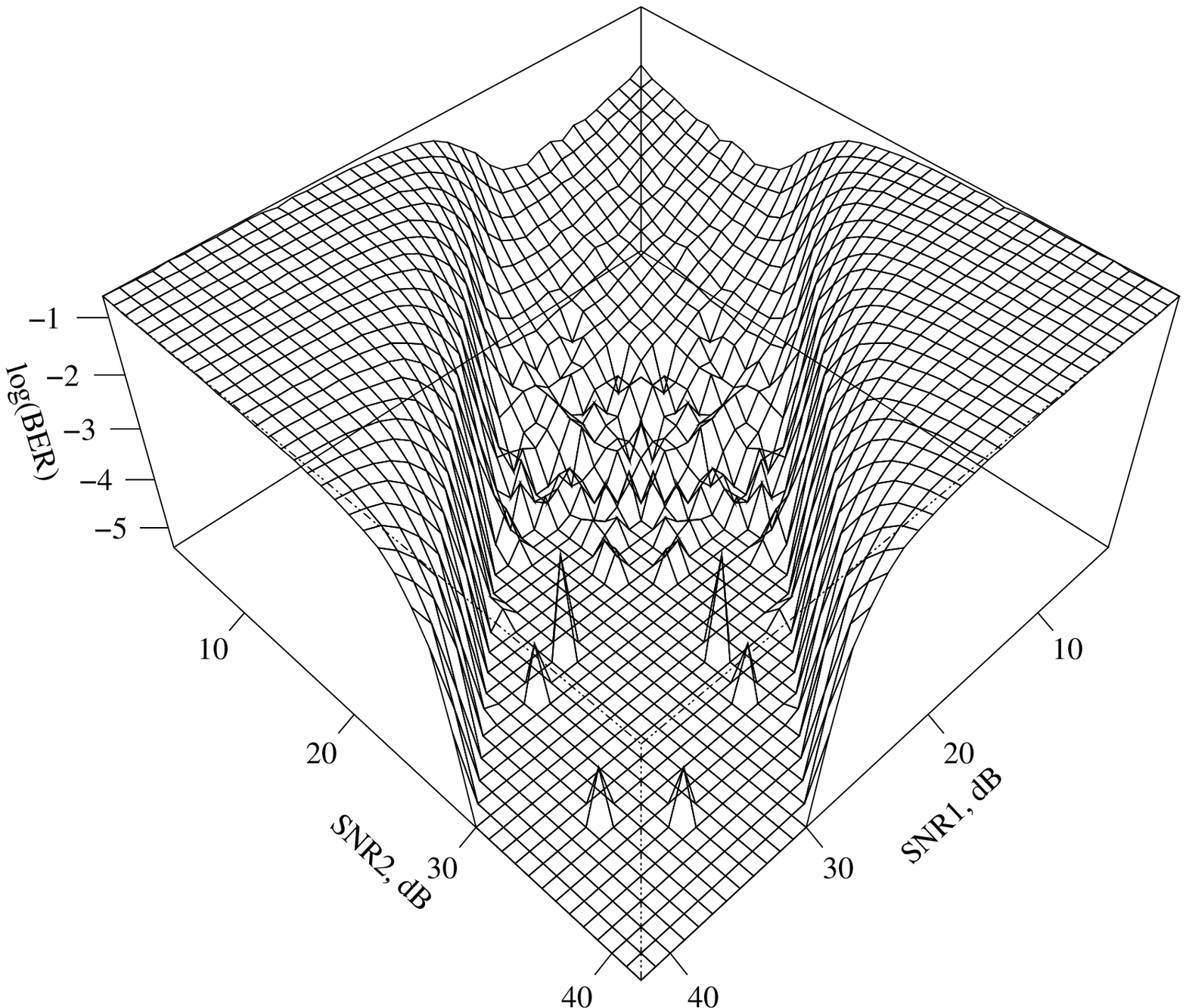}
\end{center}
\caption[BEP estimates for a space of configurations.]{(top) Estimates
of the BEPs for a space of configurations $\{c_k\}_{k=1}^M$ ($M=1600$
points at 10000 frames per point).  The $X$ and~$Y$ axes are the
average SNRs of the branches (in~dB).  The $Z$ axis is the (base ten)
logarithm of the simulated BER.  These estimates are not statistically
significant. (bottom) Statistically significant estimates
$\{\hat{b}_k\}_{k=1}^M$ of the expected BEPs $\{E[b_k]\}_{k=1}^M$ for
the same space of configurations $\{c_k\}_{k=1}^M$.  For the most part,
we are 90\% confident that the estimated expected BEP lies within 10\%
of its true value.  See text for exceptions.}
\label{fig:space}
\end{figure}

\begin{figure}
\begin{center}
\includegraphics[width=4.0in]{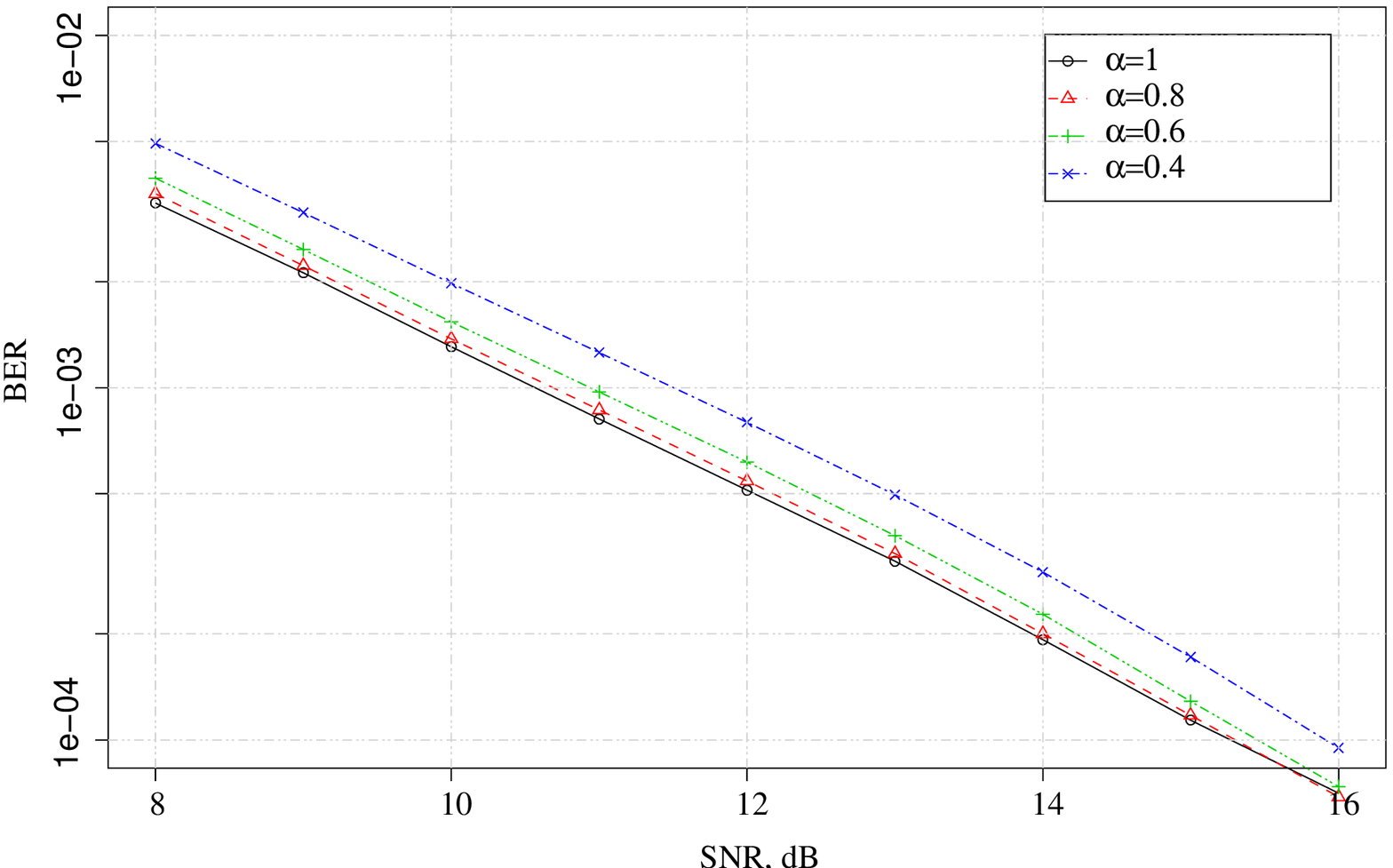} \\
\bigskip
\includegraphics[width=4.0in]{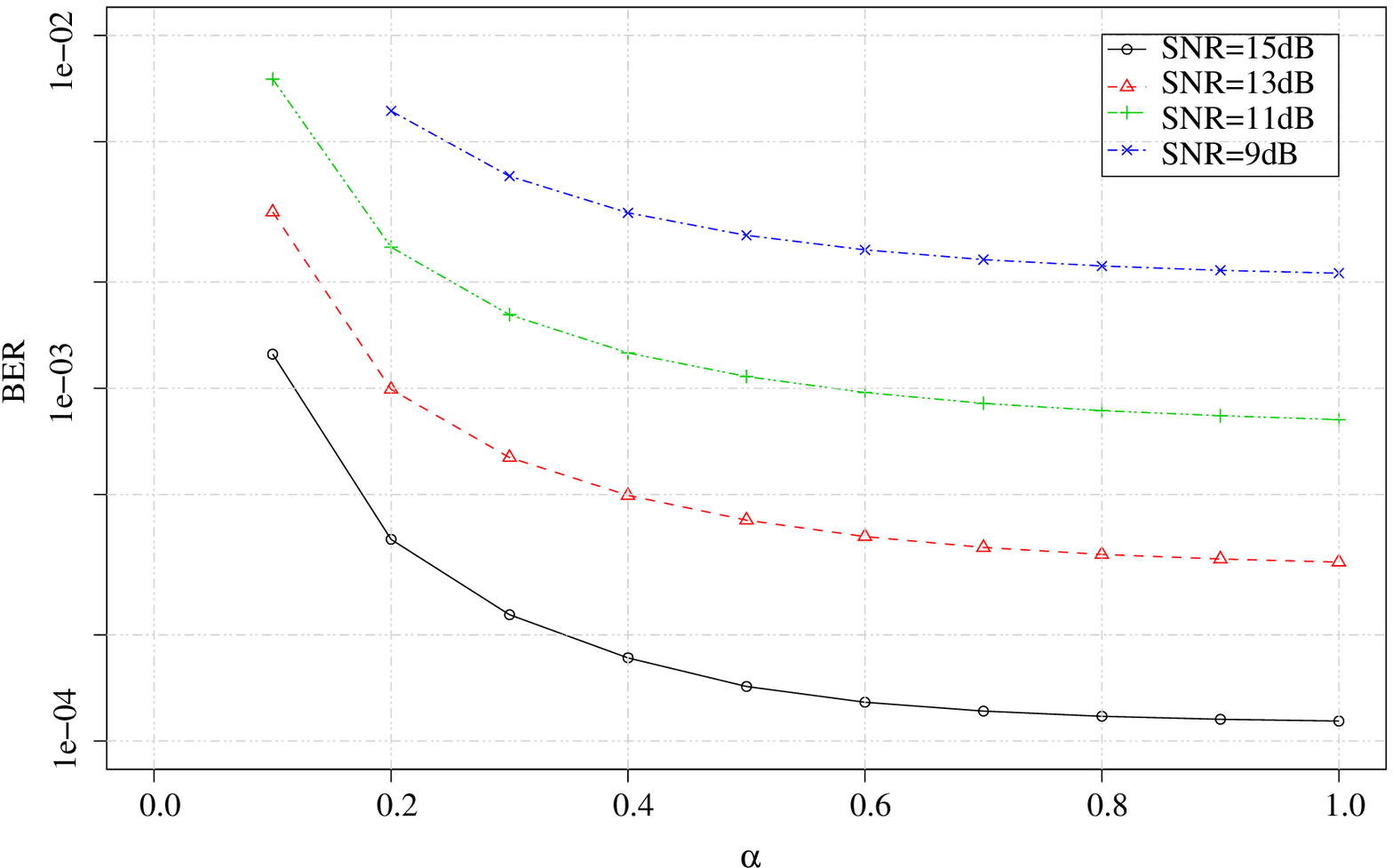}
\end{center}
\caption[1D slices of the surface in Figure~\ref{fig:space}.]{1D slices
of the configuration space $\{c_k\}_{k=1}^M$ with fixed branch power
imbalance factor $\alpha=10^{-0.1|S_1-S_2|}$ and varying effective SNR
$S=10\log_{10}\left((10^{0.1S_1}+10^{0.1S_2})/2\right)$ (top), and
fixed effective SNR~$S$ and varying branch power imbalance
factor~$\alpha$ (bottom).  These slices were computed from the surface
fit onto the data in Figure~\ref{fig:space} (bottom).  The entire
fitted surface is shown in Figure~\ref{fig:fitted}.}
\label{fig:fixed-slices}
\end{figure}

\begin{figure}
\begin{center}
\includegraphics[width=4.0in]{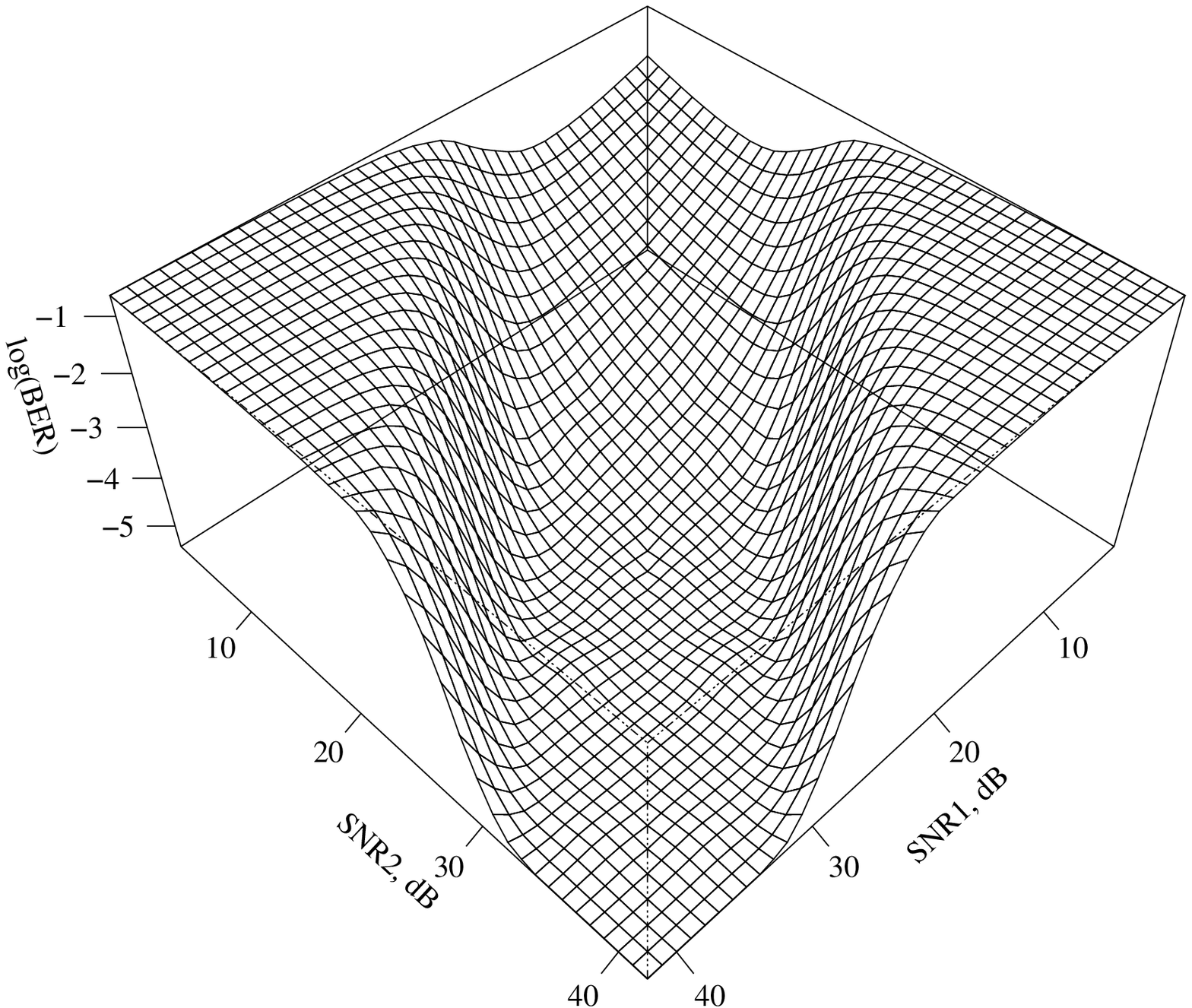}
\end{center}
\caption[A surface fitted onto the data in Figure~\ref{fig:space}.]{A
surface fitted onto the statistically significant results in
Figure~\ref{fig:space} (bottom).  We used a local linear least squares
regression with a 5\% neighborhood and tricubic weighting.  This
procedure was chosen because it can approximate the relatively steep
edge of the tolerance region.  See~\cite{dm-stat} for details.}
\label{fig:fitted}
\end{figure}

Let us begin with an initial sample of the configuration space, as
shown in Figure~\ref{fig:space} (top).  This figure shows the simulated
BER as a 2D function $\hat{f}(S_1,S_2)$ of the average branch bit
energy-to-noise ratios (SNRs) $S_1$ and~$S_2$, in~dB.  The parallel
simulation ran for three days on 120 machines (AMD Athlon 1.0~GHz) at a
speed of approximately 2.5 points per machine per day.  10000 frames,
or 800000 information bits, were simulated for each of the 820 points
$S_2=3,4,\ldots,42$; $S_1=3,4,\ldots,S_2$.  Since $\hat{f}(S_1,S_2)$ is
symmetric~\cite{sttd-stutzman}, we show $M=1600$ points
$\{c_k\}_{k=1}^M$ for a full cross-product of $S_1$ and~$S_2$.

Wireless system designers are more accustomed to 1D slices of the
configuration space, e.g., the ones shown in
Figure~\ref{fig:fixed-slices}.  Define the \emph{branch power imbalance
factor} $$\alpha=10^{-0.1|S_1-S_2|},$$ where $S_1$ and~$S_2$ are the
average SNRs of the branches, in~dB.  (This definition applies as long
as the mean path losses of the branches are equal.)  By definition,
$0\le\alpha\le{}1$, where zero corresponds to a total malfunction of
one of the branches and one corresponds to a perfect balance of branch
powers.  The graphs in Figure~\ref{fig:fixed-slices} were obtained by
fixing $\alpha$ and varying the \emph{effective SNR}
$$S=10\log_{10}\left((10^{0.1S_1}+10^{0.1S_2})/2\right),$$ in~dB (top),
and fixing the effective SNR and varying $\alpha$ (bottom).  (Note that
fixing the effective SNR is equivalent to fixing total transmitter
power.)  The sample of configurations came from the
dataset shown in Figure~\ref{fig:space} (bottom), described in detail
later. However, this sample does not contain 
the exact points for typical slices, so we
used a fitted surface---Figure~\ref{fig:fitted}---to approximate the
BERs for the slices in~Figure~\ref{fig:fixed-slices}.  We choose to
work with the axes $S_1,S_2$ in Figure~\ref{fig:space} because it
simplifies the discussion later.

What can be gathered from Figure~\ref{fig:space} (top)?  The deep
valley along the diagonal is due to the fact that, provided that the
effective SNR is fixed, we expect the BEP to be smallest when the
branch power is balanced ($S_1=S_2$, $\alpha=1$)~\cite{sttd-stutzman}.
Somewhat less expected were (a)~the wide \emph{tolerance region} where
$|S_1-S_2|$ is large (up to 12~dB) but the BER is still small, (b)~a
very sharp decline in performance at the edge of the tolerance region,
and (c)~a region of high local variability in the upper part of the
diagonal.  The surface is truncated at
$$\min_{1\le{}k\le{}M}\{\hat{b}_k\}=3.75\times{}10^{-6}$$ because
smaller estimates of the (expected) BEP require an enormous computation
time due to the convergence properties of Monte Carlo Estimation (more
on this below).

\subsection{Statistically Significant Sampling Methodology}

The initial sample looks reasonable and uncovers interesting trends in
system performance, but it does not contain enough information to make
statistically significant claims.  Estimating the probability that a
configuration exhibits acceptable average performance requires several
samples per point~$c_k$.  The simulation is computationally expensive
and different regions of the configuration space exhibit different
variability.  Therefore, we must define tight stopping criteria for
sampling.  Figure~\ref{fig:space} (bottom) shows the output obtained
with the following (per point~$c_k$) stopping criteria.  The criteria
are designed to achieve high estimation accuracy.
\begin{enumerate}
\item Sampling $\{\hat{x}_{kj}\}$ stops when the relative error in
the estimate~$\hat{b}_k$ of the expected BEP $E[b_k]$ is smaller than
the \emph{relative accuracy threshold} $\beta=0.1$ times the current
estimate~$\hat{b}_k$, at a $\gamma=0.9$ confidence level, i.e., when
$$P(|E[b_k]-\hat{b}_k|<\beta\hat{b}_k)\ge\gamma.$$ We required
$n_k\ge{}2$ samples to obtain an estimate~$\hat{\sigma}_k^2$ of the BEP
variance~$\sigma_k^2$.  Notice that the target is the relative error,
not the absolute error, because the range of $\{\hat{b}_k\}_{k=1}^M$ in
the configuration space spans four orders of magnitude.  Therefore,
absolute error measures are misleading.
\item Sampling $\{\hat{x}_{kj}\}$ also stops when we can say,
with confidence $\gamma=0.9$, that the expected BEP $E[b_k]$ is below
the \emph{sampling threshold}~$t=10^{-4}$, i.e., when
$$P(E[b_k]<t)\ge\gamma.$$ This work considers voice quality
applications, so the exact value of the expected BEP is irrelevant as
long as it is smaller than the performance threshold~$T=10^{-3}$.  The
sampling threshold~$t$ was set to an order of magnitude below the
performance threshold~$T$ to avoid large approximation error of a
fitted surface near~$T$.
\item Finally, sampling $\{\hat{x}_{kj}\}$ stops when more than
50 samples of 10000 frames each are required to satisfy either of the
previous rules.  This rule fired in 5\% of the cases, all at the
boundary of the tolerance region and most in mid diagonal.
\end{enumerate}

\noindent Altogether, 5154 samples were collected for an average of 6.3
samples per point.  Needless to say, the computational expense of such
sampling remains too high for practical applications. While a large
number of samples is typically desirable (for validation purposes),
we will show that our data mining
framework makes very effective use of data 
and thus requires fewer samples in practice.
Let us now look at
the data in more detail.

\subsection{Results of Statistically Significant Sampling}

\begin{figure}
\begin{center}
\includegraphics[width=4.0in]{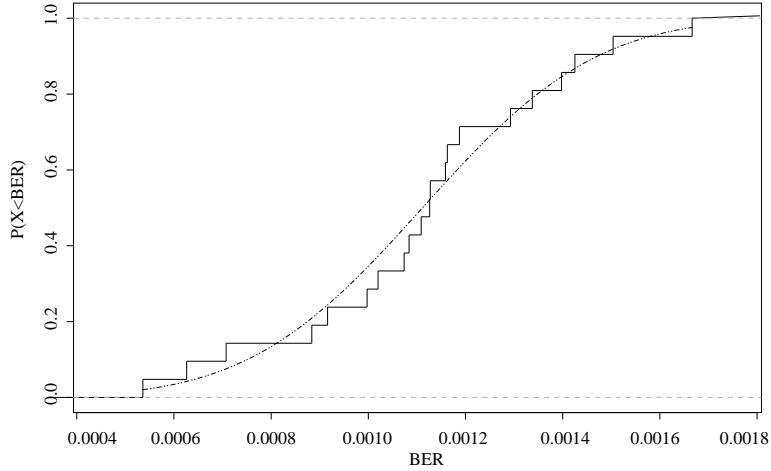}
\end{center}
\caption[Empirical CDF for one of the configurations.]{Empirical CDF of
21 samples for a randomly chosen point vs. that of the Gaussian
distribution with appropriate mean and variance.}
\label{fig:ecdf}
\end{figure}

It is also likely that the samples output by the WCDMA simulation are
approximately Gaussian distributed. Intuitively,
we are simulating a large number of information bits (800000) per BEP
estimate $\hat{x}_{kj}$, so the Lindeberg condition for the Central
Limit Theorem should hold.  Figure~\ref{fig:ecdf} shows empirical
evidence that this is the case.  We have arbitrarily chosen one point
among those with 20--30 sample values $\{\hat{x}_{kj}\}_{j=1}^{n_k}$
and plotted the empirical CDF of this sample against that of the
Gaussian distribution with the mean equal to sample mean~$\hat{b}_k$
and the variance equal to sample variance~$\hat{\sigma}_k^2$.  The
curves are close to each other and the Shapiro-Wilk test yields
$W=0.98$ ($0\le W\le 1$) and $p$-value of~$0.88$.  Other points also
demonstrate similar curves and high values of~$W$, but $p$-values vary
significantly.  This dataset contains sufficient samples to estimate
$\{E[b_k]\}_{k=1}^M$ with high relative accuracy, but 6.3 samples per
point are insufficient to formally justify a Gaussian assumption.

\begin{figure}
\begin{center}
\includegraphics[width=4.0in]{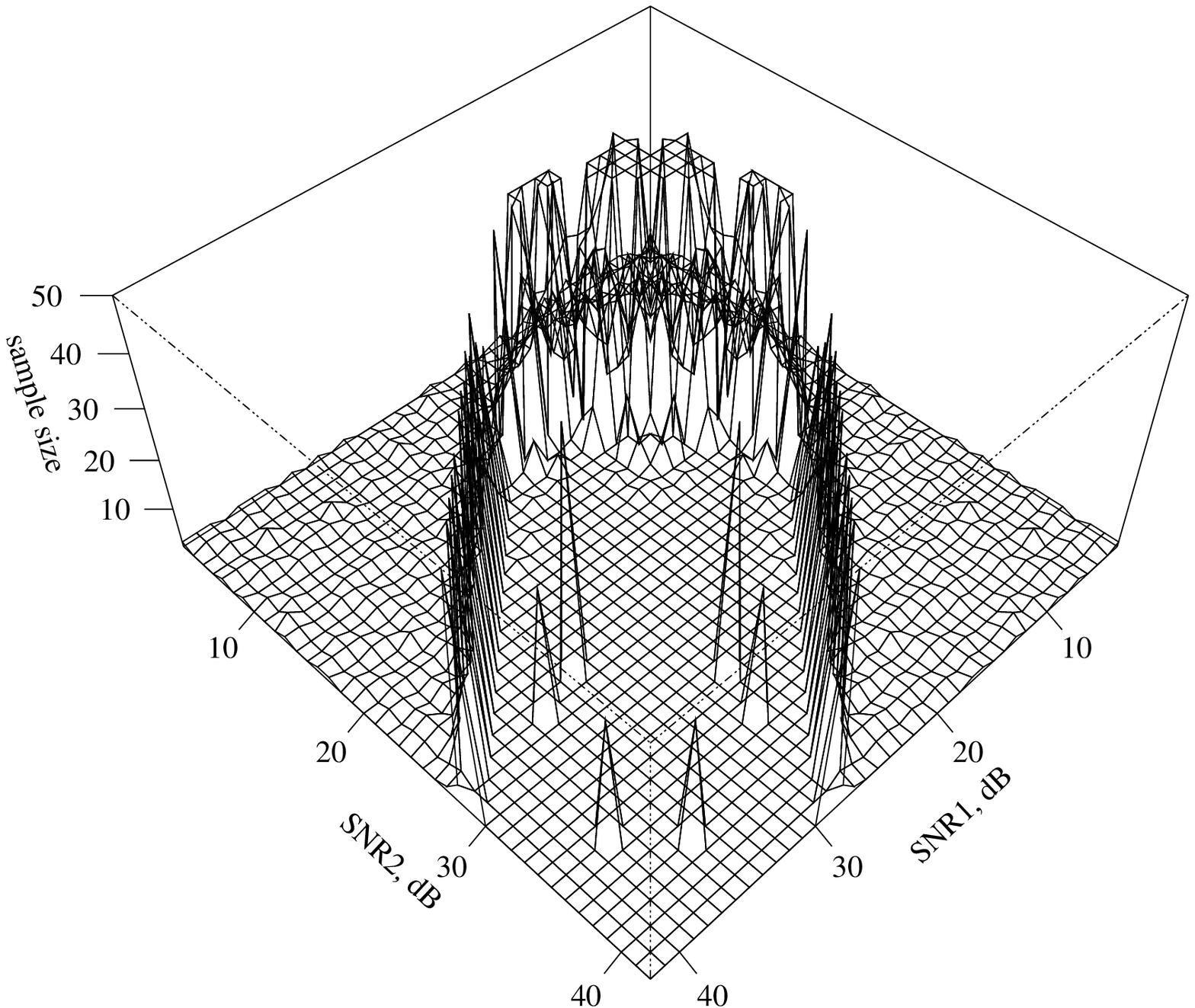} \\
\bigskip
\includegraphics[width=4.0in]{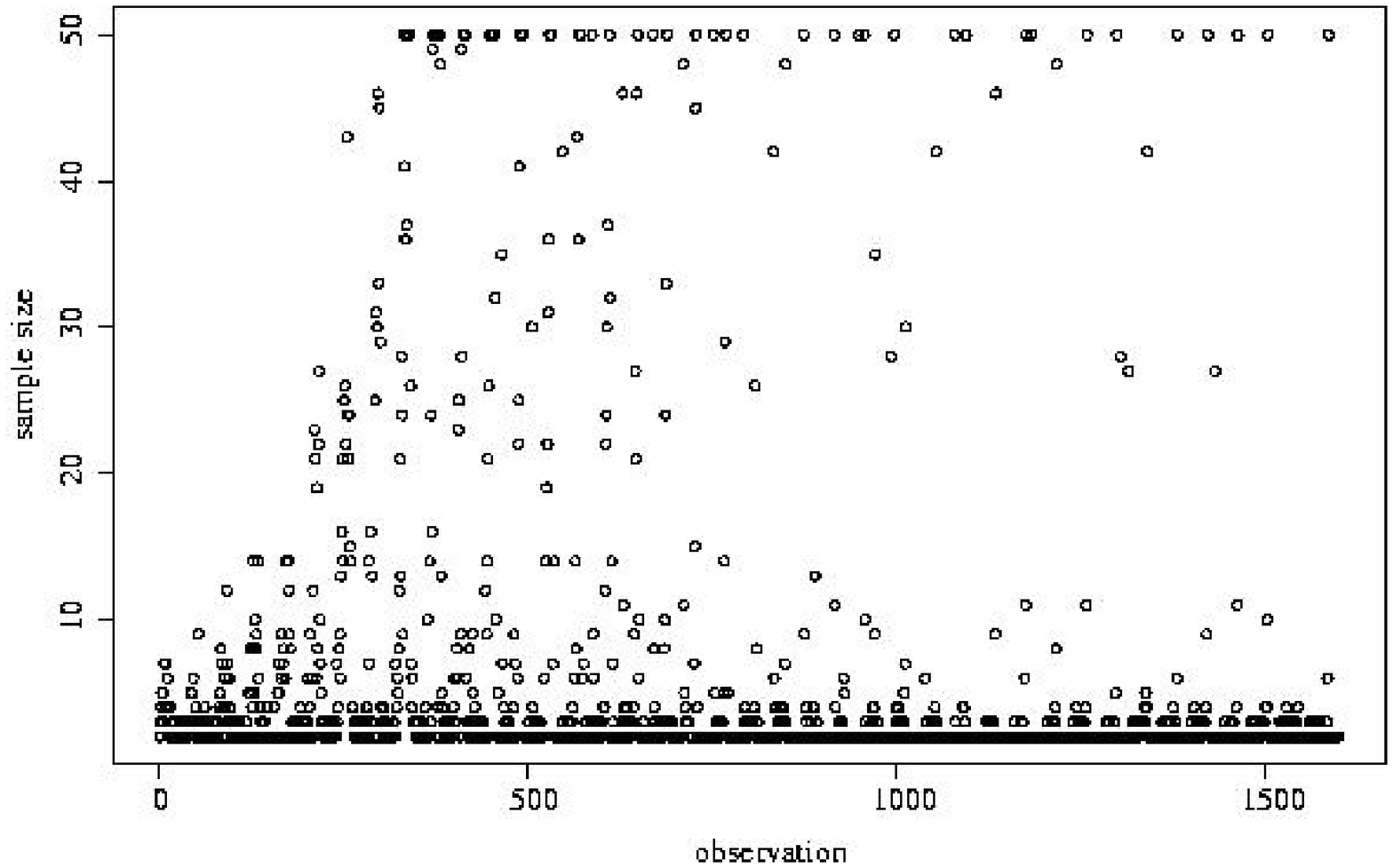}
\end{center}
\caption[Sample sizes for Figure~\ref{fig:space}.]{Sample sizes for
Figure~\ref{fig:space} (bottom).  The top part shows the perspective
plot and the bottom part shows the scatter plot.}
\label{fig:sample-size}
\end{figure}

\begin{figure}
\begin{center}
\includegraphics[width=4.0in]{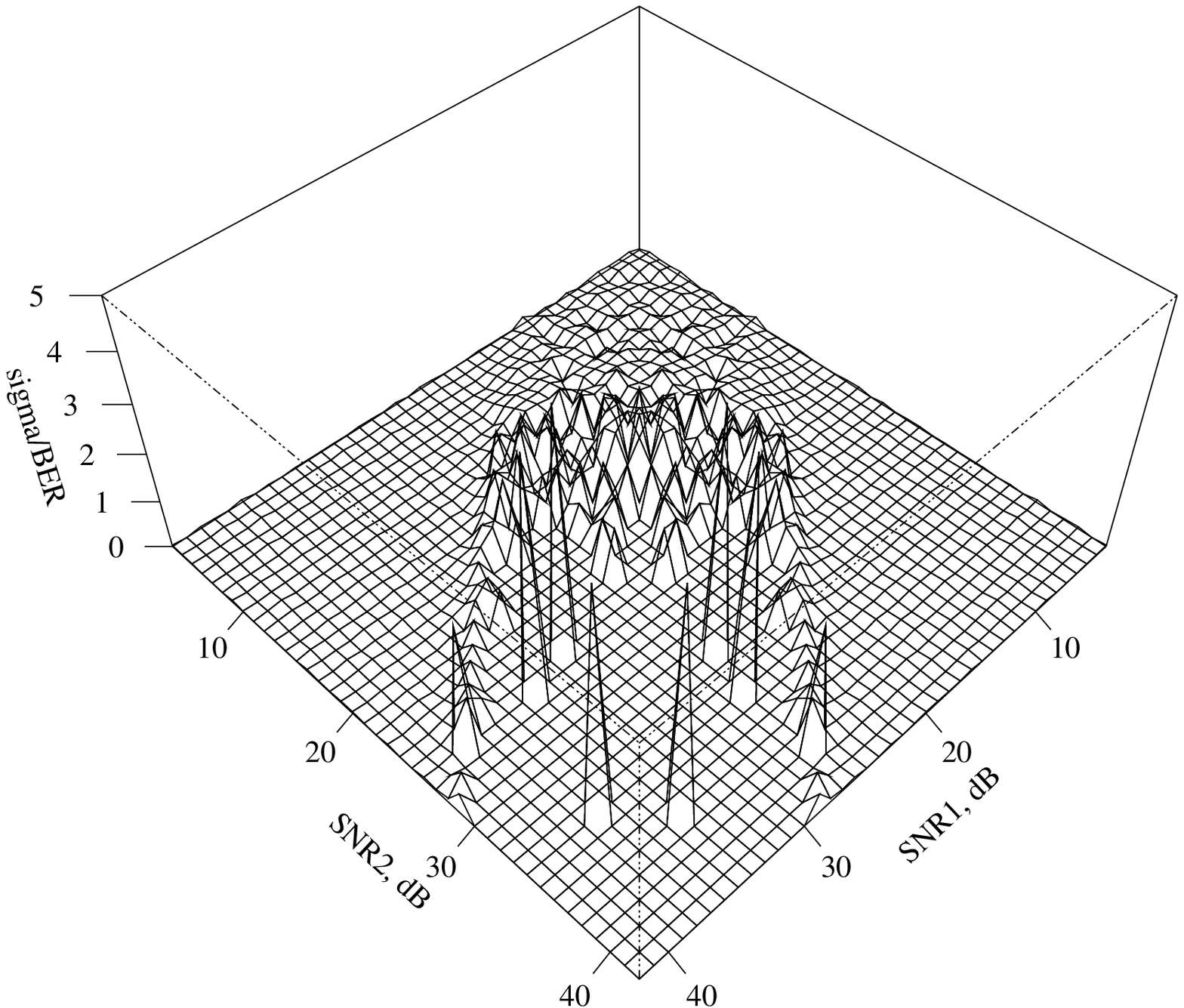} \\
\bigskip
\includegraphics[width=4.0in]{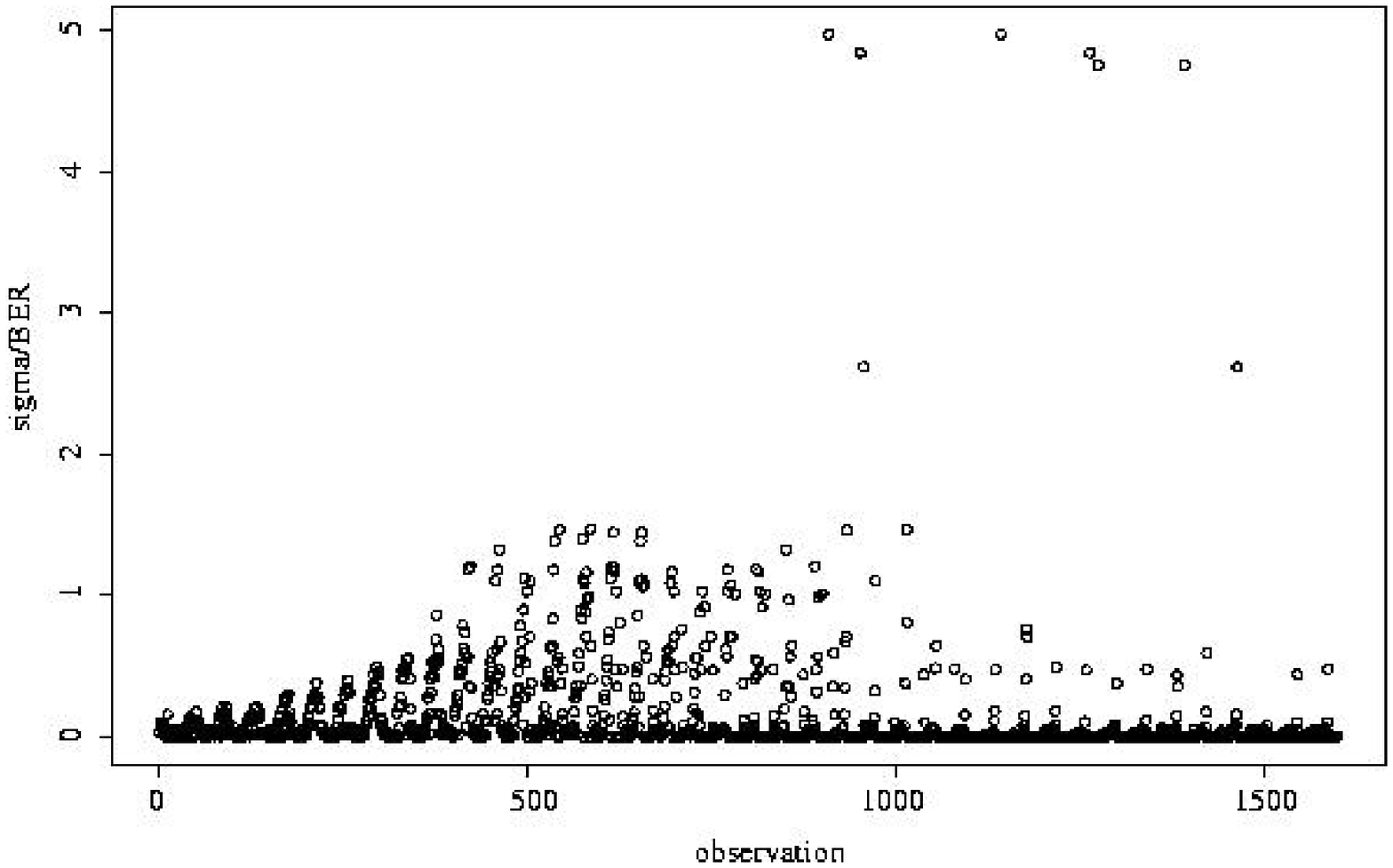}
\end{center}
\caption[Sample standard-deviation-to-mean ratios for
Figure~\ref{fig:space}.]{Sample standard deviation-to-mean ratios for
Figure~\ref{fig:space} (bottom).  The top part shows the perspective
plot and the bottom part shows the scatter plot.}
\label{fig:sample-error}
\end{figure}

It is also instructive to see some measure of how the sample variance
is distributed across the configuration space.
Figures~\ref{fig:sample-size} and~\ref{fig:sample-error} show sample
sizes and sample standard deviation-to-mean ratios for the samples in
Figure~\ref{fig:space} (recall that we prefer relative measures because
the range of $\{\hat{b}_k\}_{k=1}^M$ is large).  Both figures indicate
high variance around the boundary of the tolerance region.  This is not
surprising because the edges of the tolerance region are relatively
steep.  Figure~\ref{fig:sample-error} also shows relatively high
variance at some points inside the tolerance region.  This is because
the simulation achieved the sampling threshold~$t=10^{-4}$ and stopped
before it achieved the relative accuracy threshold~$\beta=0.1$.
Knowing this, one would expect a larger relative variance in the
tolerance region.  Let us examine why this is not the case.

We treat the BEP as a continuous Gaussian random variable~$b_k$, but
all sample values $\{\hat{x}_{kj}\}_{j=1}^{n_k}$ are discrete---they
are ratios of two integers, the number of errors and the number of bits
simulated.  The simulation may not detect any bit errors when the
expected BEP $E[b_k]$ is relatively small (e.g., one error in the
number of bits simulated).  Since no channel is perfect, zero is too
optimistic an estimate for the expected BEP.  Instead, we
conservatively assume that at least three bit errors have been
detected.  This is why the smallest estimate~$\hat{b}_k$ of~$E[b_k]$ is
$3/800000=3.75\times{}10^{-6}$.  However, using any constant cutoff
prevents us from estimating the variance $\sigma_k^2$.  We would need
to simulate a large number of frames to estimate $\sigma_k^2$ when the
expected BEP is small.  Instead, we can empirically show that the
probability that the expected BEP is smaller than the performance
threshold~$T=10^{-3}$ is close to one.  Let
$\hat{b}_k=3.75\times{}10^{-6}$ be the sample mean, $n_k=2$ be the
sample size, and $\sigma_k^2$ be the BEP variance at point~$c_k$ where
two independent simulations detected three or fewer bit errors each.
Sampling $\{\hat{x}_{kj}\}$ will stop because sample variance is zero,
so the first stopping rule applies.

We need to show that sampling can indeed stop, i.e., that the
probability that the expected BEP is below the performance
threshold~$T$ is
$$P(E[b_k]<T)\approx{}F_{n_k-1}\left(\frac{T-\hat{b}_k}{\hat{\sigma}_k/\sqrt{n_k}}\right)\ge{}0.995.$$
This statement can only be false when
$(T-\hat{b}_k)\sqrt{n_k}/\hat{\sigma}_k\le{}64$, or
$\hat{\sigma}_k\ge{}2.2\times{}10^{-5}$, almost an order of magnitude bigger
than the conservative estimate $\hat{b}_k$ of the expected BEP
$E[b_k]$.  This is unlikely because Figure~\ref{fig:sample-error}
(bottom) shows that the sample standard deviation rarely exceeds the
sample mean even by half an order of magnitude.  In other words, we do
not have accurate estimates for variance~$\sigma_k^2$ in the tolerance
region.  However, we can still reasonably conclude that configurations
exhibit acceptable performance in this region.


\section{The Third Level of Aggregation: Regions}
\label{sec:gizmo}

Consider a set of buckets $\{C_k\}_{k=1}^M$ with corresponding random
variables $\{B_k\}_{k=1}^M$.  Given a number of sample values, the
framework developed in Section~\ref{sec:stat} allows us to estimate the
probabilities $\{P(E[B_k]<T)\}_{k=1}^M$ that buckets $\{C_k\}_{k=1}^M$
exhibit acceptable average performance.  (All arguments about buckets
equally apply to points because a point is a special case of a bucket.)
This section is concerned with finding an optimal subset of random
variables from among $\{B_k\}_{k=1}^M$.  This optimal subset
corresponds to an optimal region of a 2D bucket space.  We would like
to find a sufficiently large admissible region~$R_m$ such that we are
sufficiently confident that buckets in~$R_m$ exhibit acceptable average
performance.

There are many ways to define admissibility and we are interested in
adopting a definition that is both meaningful in the wireless domain
and permits effective data mining algorithms. Among a space of such
admissible regions, we can define different optimality criteria and
data mining then reduces to searching within this space. In this
\paper{}, a region~$R_m$ is admissible when it has a particular type of
shape.  We will explore three different criteria for the mining of
optimal regions; the algorithms and these criteria are based on the
work of Fukuda et al.~\cite{fukuda-rectilinear} and have been adapted
to the problem of mining simulation data in this \paper{}.

\begin{table}
\begin{center}
\begin{tabular}{c l}
\hline
$X,Y$ & parameters that partition the point space into buckets \\
$M_X,M_Y$ & $X$ and $Y$ dimensions of the bucket space \\
$M=M_X\times{}M_Y$ & number of buckets in the bucket space \\
$D_X,D_Y$ & domains of $X$ and $Y$ \\
$\eta(m)$ & number of buckets in region $R_m$ \\
$C_{\kappa(m,i)}$ & $i$-th bucket in region $R_m$, $1\le{}i\le{}\eta(m)$ \\
$n_{\kappa(m,i)}$ & number of samples in bucket $C_{\kappa(m,i)}$\\
$x_{\kappa(m,i)},y_{\kappa(m,i)}$ & $X$ and $Y$ values for bucket $C_{\kappa(m,i)}$ \\
\hline
\end{tabular}
\end{center}
\caption[Summary of region notation.]{Summary of region notation.  Also
see Table~\ref{tab:notation}.}
\label{tab:notation2}
\end{table}

Additional notation relating buckets to regions is introduced in
Table~\ref{tab:notation2}.  Let $X$ and~$Y$ be two discrete parameters
to the temporal (e.g., WCDMA) simulations such that $X$ and~$Y$
partition the point space into disjoint buckets $\{C_k\}_{k=1}^M$.
More precisely, let $X,Y$ have ordinal domains $D_X,D_Y$, let
$|D_X|=M_X,|D_Y|=M_Y,|D_X||D_Y|=M$, and assume that the map
$\rho:D_X\times{}D_Y\rightarrow{}\{C_k\}_{k=1}^M$ is bijective.  In
other words, $X$ and~$Y$ define a discrete 2D space of buckets.  Since
the domains of $X$ and~$Y$ are ordinal, this space is easily visualized
as a 2D color map or a 3D perspective plot.

\begin{figure}
\begin{center}
\includegraphics[width=4.0in]{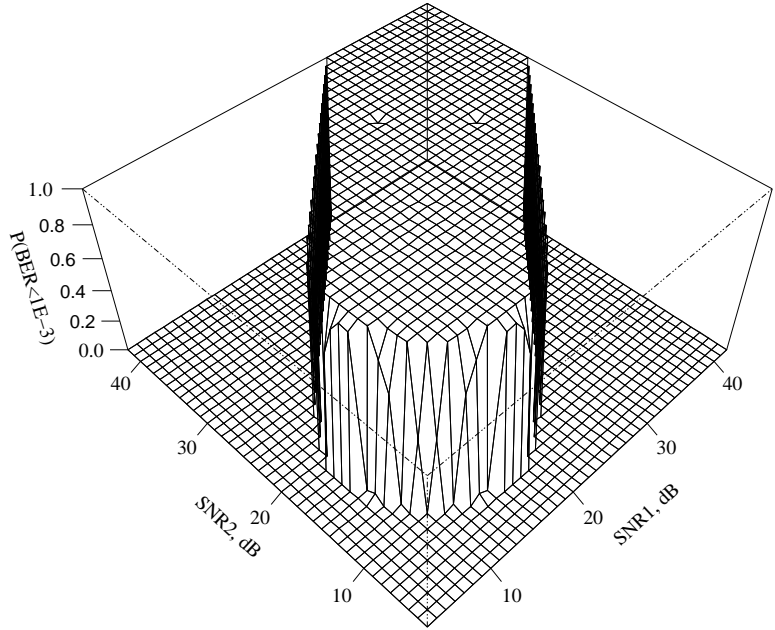}
\end{center}
\caption[Probabilities that configurations in Figure~\ref{fig:space}
exhibit acceptable performance.]{Probabilities
$\{P(E[b_k]<T)\}_{k=1}^M$ that configurations $\{c_k\}_{k=1}^M$
exhibit acceptable performance with respect to the performance
threshold~$T=10^{-3}$ (voice quality system).  This perspective plot
corresponds to the STTD dataset in Figure~\ref{fig:space} (bottom).
The axes $S_1$ and~$S_2$ are rotated 180 degrees counter-clockwise to
provide a better view of the surface.}
\label{fig:probabilities}
\end{figure}

For example, the average SNRs $S_1$ and~$S_2$ in the previous section
partition the space of configurations into buckets.  Both $S_1$
and~$S_2$ vary from~3 to~42 in steps of~1 (in~dB), so $M_X=M_Y=40$ and
$M=40\times{}40=1600$ (recall, from Section~\ref{sec:w-example}, that
only 820 of these points were simulated and the remaining ones were
symmetrically reflected).  Furthermore, the domains of $S_1$ and~$S_2$
are ordinal because the values of~$S_1$ and~$S_2$ are directly related
to the powers of the transmitter antennas.  In this case, the buckets
are simply the points in the space of configurations.  In general,
buckets can be convex combinations of points, as detailed in
Section~\ref{sec:stat}.  Recall that we defined the color of a bucket
as the probability that the bucket exhibits acceptable average
performance.  Figure~\ref{fig:probabilities} shows these `colors' as a
perspective plot for the STTD example.

\subsection{Region Shape}

Consider regions (subsets) of buckets in the bucket space.  If the
shape of these regions is unconstrained, there are $2^M$ possible
regions $\{R_m\}_{m=1}^{2^M}$.  Let region $R_m$, $1\le{}m\le{}2^M$,
consist of buckets $\{C_{\kappa(m,i)}\}_{i=1}^{\eta(m)}$, where
$\eta(m)$, $1\le{}m\le{}2^M$, is a mapping from region number~$m$ to
the number of buckets in this region, and $\kappa(m,i)$,
$1\le{}m\le{}2^M$, $1\le{}i\le\eta(m)$, is a mapping from region
number~$m$ and bucket number~$i$ within region~$R_m$ to bucket
number~$k$, $1\le{}k\le{}M$, that we use to subscript buckets
$\{C_k\}_{k=1}^M$.  The exact definitions of $\eta(m)$ and
$\kappa(m,i)$ are not important as long as they generate all possible
regions (subsets) $\{R_m\}_{m=1}^{2^M}$.

\begin{figure}
\begin{center}
\includegraphics{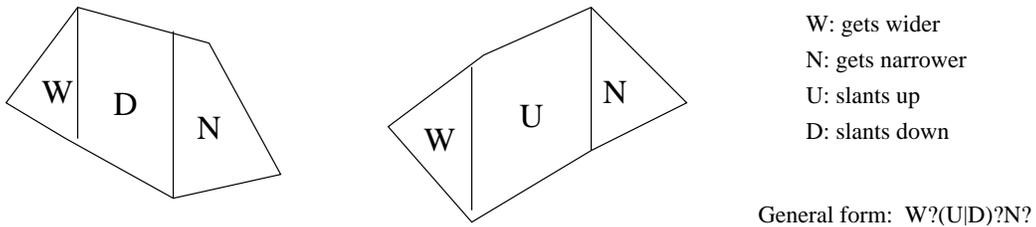}
\end{center}
\caption[Types of admissible regions.]{Some types of admissible
(connected rectilinear) regions.  When we look at an admissible region
from left to right, its upper boundary must first increase and then
decrease monotonically, and its lower boundary must first decrease and
then increase monotonically.}
\label{fig:admissible}
\end{figure}

The shape of admissible regions should be constrained because
unconstrained regions are hard to interpret and tend to overfit the
training data.  Besides, the problem of selecting an optimal
unconstrained region is computationally intractable---all $2^M$
possible regions must be considered, where $M=1600$ in the STTD
example.  The region shape can be constrained in a number of different
ways (rectangular, x-monotone, etc.).  Our restrictions on region shape
are discussed next.

Without loss of generality, assume that $D_X=\{1,2,\ldots,M_X\}$ and
$D_Y=\{1,2,\ldots,M_Y\}$.  Intuitively, region~$R_m$ is rectilinear
when its intersection with any horizontal or vertical line is
connected.  More formally, region~$R_m$ is \emph{rectilinear} if and
only if whenever buckets $C_{\kappa(m,i)}$ at
$(x_{\kappa(m,i)},y_{\kappa(m,i)})$ and $C_{\kappa(m,j)}$ at
$(x_{\kappa(m,j)},y_{\kappa(m,j)})$ are both in~$R_m$, then
(a)~$\rho(r,s) = C_{\kappa(m,i)}$ and $\rho(r,t) = C_{\kappa(m,j)}$ 
imply buckets $\rho(r,u)$ are also in~$R_m$ for all
$u \in [s,t]$, and
(b)~$\rho(r,t) = C_{\kappa(m,i)}$ and $\rho(s,t) = C_{\kappa(m,j)}$ 
imply buckets $\rho(u,t)$ are also in~$R_m$ for all
$u \in [r,s]$. Here $[a,b]$ means all integers between the integers
$a$, $b$, inclusive.
%
We use Manhattan geometry to define
connectedness.  Region~$R_m$ is \emph{connected} if and only if for
every pair of buckets $C_{\kappa(m,i)}$ and $C_{\kappa(m,j)}$ in~$R_m$
there exists a sequence of buckets
$$C_{\kappa(m,i)}=C_{\kappa(m,l_1)},C_{\kappa(m,l_2)},\ldots,C_{\kappa(m,l_n)}=C_{\kappa(m,j)}$$
in~$R_m$ such that for every $1\le{}k<n$
$${\parallel \rho^{-1} (C_{\kappa(m,l_k)}) - \rho^{-1} (C_{\kappa(m,l_{k+1})}) \parallel}_1 = 1.$$
Furthermore, we say that region~$R_m$ is \emph{admissible} if it is
both rectilinear and connected.

This definition of admissibility can be viewed as a relaxed definition
of convexity.  Geometrically, it is easy to see that region~$R_m$ is
admissible if and only if, when we look at~$R_m$ from left to right,
its upper boundary first increases and then decreases monotonically (a
pseudoconcave function), and its lower boundary first decreases and
then increases monotonically (a pseudoconvex function).  In other
words, the region boundary need not be strictly convex or strictly
concave, but it must be pseudoconvex or pseudoconcave.  Admissible
regions are informally summarized in Figure~\ref{fig:admissible}.  All
admissible regions are composed of regions of four primitive types: W
(region gets wider from left to right), N (region gets narrower), U
(region slants up), and D (region slants down).  Twelve combinations of
these types yield all types of admissible regions: W, WU, WUN, WD, WDN,
WN, UN, DN, U, D, N, and the empty region.

Our choice of connected rectilinear regions is due to primarily
heuristic considerations.  These considerations are commonly
applicable, but must be re-evaluated for each study.  Both the
connectedness and the rectilinearity restrictions can be justified for
the STTD example (see next section).  In general, it is easy to justify
connectedness, but hard to justify rectilinearity.  We advocate the use
of connected rectilinear regions primarily because this shape is
resistant to noise in the sample, not because we can analytically show
that the region boundary is rectilinear.  In data mining, the choice of
region shape is most commonly dictated by the desired tradeoff between
bias and variance~\cite{dm-stat}.  Regions with flexible shape exhibit
small bias (they can fit any data) but high variance (they can be
overly sensitive to a particular dataset).  Regions with rigid shape
exhibit high bias but small variance.  Connected rectilinear regions
provide a reasonable tradeoff between bias and variance for many
applications.

\subsection{Evaluating Regions}

Another prerequisite to finding regions with the desired properties is
a definition of region `goodness'.  Let us map bucket confidence
$P(E[B_{\kappa(m,i)}]<T)$ to a discrete range $[0\ldots{}1000]$ and
define the \emph{hit of bucket} $C_{\kappa(m,i)}$ as
$$h_{\kappa(m,i)}=\lfloor{}1000P(E[B_{\kappa(m,i)}]<T)+0.5\rfloor$$
($\lfloor{}X\rfloor$ denotes the largest integer that does not exceed
$X$), the \emph{support of bucket} $C_{\kappa(m,i)}$ as
$$s_{\kappa(m,i)}=1000$$ (this constant was chosen to make the
discretization error reasonably small), the \emph{hit of region} $R_m$
as $$H_m=\sum_{i=1}^{\eta(m)}h_{\kappa(m,i)},$$ and the \emph{support
of region} $R_m$ as
$$S_m=\sum_{i=1}^{\eta(m)}s_{\kappa(m,i)}=1000\eta(m).$$  The key to
efficient computation of optimized-confidence and optimized-support
admissible regions is the definition of region confidence as
$$\Theta_m=H_m/S_m,$$ where $H_m$ is the hit and~$S_m$ is the support
of region~$R_m$.  Let us explore the implications of these definitions
in more detail.

\subsubsection{Model-Based and Model-Free Analyses}

Suppose, $n_{\kappa(m,i)}=6$ samples have been collected for bucket
$C_{\kappa(m,i)}$ that consists of a single point.  Let the sample mean
be $\hat{B}_{\kappa(m,i)}=5\times{}10^{-4}$ and the sample standard
deviation be $\hat{\Sigma}_{\kappa(m,i)}=8.87\times{}10^{-4}$.
Furthermore, suppose that five of these samples have the BER below
$10^{-3}$ and one has the BER above $10^{-3}$.  Then,
$$P(E[B_{\kappa(m,i)}]<T)\approx{}F_5\left(\frac{10^{-3}-5\times{}10^{-4}}{8.87\times{}10^{-4}/\sqrt{6}}\right)\approx{}0.887.$$
A purely model-free approach would interpret the above simulation
results as `bucket $C_{\kappa(m,i)}$ will exhibit acceptable average
performance in~5 out of~6 cases.' A strongly model-based approach would
interpret the simulation results as `we are 88.7\% confident that
bucket $C_{\kappa(m,i)}$ exhibits acceptable average performance.' Our
interpretation lies between the model-based approach and a model-free
approach and posits that `bucket $C_{\kappa(m,i)}$ will exhibit
acceptable average performance in~887 out of~1000 cases.' These
interpretations provide confidence estimates under different
simplifying assumptions.

The model-free interpretation does not take either sample variance or
sample distribution into account.  This interpretation is only reliable
for a sufficiently large number of samples, which is a luxury in our
application.  Our middle-ground interpretation explicitly accounts for
sample variance and sample distribution.  When sample size is small,
our interpretation provides a statistically valid estimate of
confidence that the bucket exhibits acceptable average performance.
For a single bucket, this interpretation is as good as a strongly
model-based interpretation, modulo a reasonably small discretization
error.  However, our interpretation diverges from the model-based
interpretation at the region level.

A strongly model-based analysis procedure would define a region random
variable
$$Q_m=\frac{1}{W_m}\sum_{i=1}^{\eta(m)}w_{\kappa(m,i)}B_{\kappa(m,i)},$$
where $\{B_{\kappa(m,i)}\}_{i=1}^{\eta(m)}$ are bucket random
variables, $\{w_{\kappa(m,i)}\}_{i=1}^{\eta(m)}$ are \emph{a priori}
(positive) constant weights, and
$$W_m=\sum_{i=1}^{\eta(m)}w_{\kappa(m,i)}$$ is a normalization factor
that maps these weights to probabilities of bucket occurrence in the
region.  A procedure similar to that in Section~\ref{sec:stat} would
then be used to estimate $P(E[Q_m]<T)$ for a threshold~$T$.  The
result of this calculation can be interpreted as the probability that
region~$R_m$ exhibits acceptable average performance, conditional on
the temporal simulation assumptions, the bucketing prior probabilities,
and the region prior probabilities.  However, as we shall see later,
this definition of region confidence violates a property that permits
an efficient data mining algorithm.

We think of region confidence in terms of average bucket confidence
over the whole region, namely,
$$\Theta_m\approx{}\frac{1}{\eta(m)}\sum_{i=1}^{\eta(m)}P(E[B_{\kappa(m,i)}]<T).$$
(If region size $\eta(m)$ is large enough, we can reasonably expect the
discretization errors to cancel each other.) This interpretation
of~$\Theta_m$ does not correspond to the strongly model-based
probability that region~$R_m$ exhibits acceptable average performance.
Instead, we define a region random variable~$P_m$ as the probability
that \emph{any} bucket $C_{\kappa(m,i)}$ in region~$R_m$ exhibits
acceptable average performance.  Then, we estimate the expected value
$E[P_m]$ across the region~$R_m$ by the sample mean
$\hat{P}_m\approx\Theta_m$ of estimates of bucket confidences
$\{P(E[B_{\kappa(m,i)}]<T)\}_{i=1}^{\eta(m)}$.

How do these two definitions relate to each other?  It is easy to show
that they are equivalent only under very restrictive assumptions.
Basically, we are assuming that the buckets are mutually independent,
that population variance is small, and that the region is consistent,
i.e., `good' and `bad' buckets are never mixed in the same region.  Let
bucket random variables $\{B_{\kappa(m,i)}\}_{i=1}^{\eta(m)}$ be
mutually independent, let the estimates
$\{\hat{\Sigma}_{\kappa(m,i)}^2\}_{i=1}^{\eta(m)}$ of bucket variances
be approximately equal to zero, and let the estimates
$\{\hat{B}_{\kappa(m,i)}\}_{i=1}^{\eta(m)}$ of bucket expected BEPs be
either all greater than the performance threshold~$T$ or all smaller
than the performance threshold~$T$ (i.e., all
$\{T-\hat{B}_{\kappa(m,i)}\}_{i=1}^{\eta(m)}$ have the same sign).
Then, bucket confidences
$$P(E[B_{\kappa(m,i)}]<T)\approx{}F_{n_{\kappa(m,i)}-1}\left(\frac{T-\hat{B}_{\kappa(m,i)}}{\hat{\Sigma}_{\kappa(m,i)}/\sqrt{n_{\kappa(m,i)}}}\right),$$
$1\le{}i\le\eta(m)$, will be either all approximately equal to zero
($\hat{B}_{\kappa(m,i)}>T$), or all approximately equal to one
($\hat{B}_{\kappa(m,i)}<T$).  Therefore, region confidence~$\Theta_m$
will be approximately equal to zero or one.  Likewise, the strongly
model-based region confidence
$$P(E[Q_m]<T)\approx{}F_{\eta(m)-1}\left(\frac{T-\hat{Q}_m}{\hat{\Psi}_m/\sqrt{\eta(m)}}\right)$$
will be approximately equal to zero or one because the estimate
$\hat{\Psi}^2_m$ of region variance is (see Section~\ref{sec:stat})
$$\hat{\Psi}_m^2=\frac{1}{W_m^2}\sum_{i=1}^{\eta(m)}w_{\kappa(m,i)}^2\hat{\Sigma}_{\kappa(m,i)}^2\approx{}0.$$
The sign of $T-\hat{Q}_m$ determines whether $P(E[Q_m]<T)$ is
approximately equal to zero or one.  After a minor rearrangement of
terms,
$$T-\hat{Q}_m=\frac{1}{W_m}\sum_{i=1}^{\eta(m)}w_{\kappa(m,i)}(T-\hat{B}_{\kappa(m,i)}).$$
We assumed that $\{T-\hat{B}_{\kappa(m,i)}\}_{i=1}^{\eta(m)}$ have the
same sign, so we have shown that $P(E[Q_m]<T)\approx\Theta_m$.  The
equality is asymptotically exact as all variance estimates
$\{\hat{\Sigma}_{\kappa(m,i)}^2\}_{i=1}^{\eta(m)}$ approach zero.  This
argument applies regardless of the distributions of
$\{B_{\kappa(m,i)}\}_{i=1}^{\eta(m)}$, as long as these random
variables are mutually independent.

\subsection{Optimized Regions}

We now pursue the definition of optimized regions.  Given a
{\it slope}~$\tau$, $0\le\tau\le{}1$, define the \emph{gain} of region~$R_m$,
$1\le{}m\le{}2^M$, as $$G(R_m,\tau)=H_m-\tau{}S_m,$$ where $H_m$ is the
region hit and~$S_m$ is the region support.  Let an
\emph{optimized-gain admissible region}~$R_\tau$ with respect to
slope~$\tau$, $0\le{}\tau\le{}1$, be an admissible region with the
maximum gain $G(R_\tau,\tau)$ over all admissible regions (this region
need not be unique).  Optimized-gain admissible regions are easy to
define, compute, and analyze, but hard to interpret.  Common practice
is to define optimized-confidence and optimized-support admissible
regions.  Admissible region~$R_*$ is an \emph{optimized-confidence
admissible region} with respect to a given support
threshold~$1000\eta$, $0\le{}\eta\le{}M$, if $R_*$ has the maximum
confidence $\Theta_*=H_*/S_*$ among all admissible regions with support
of at least $1000\eta$.  Likewise, admissible region~$R_\diamond$ is an
\emph{optimized-support admissible region} with respect to a given
confidence threshold~$\theta$, $0\le{}\theta\le{}1$, if $R_\diamond$
has the maximum support $S_\diamond=1000\eta(\diamond)$ among all
admissible regions with confidence of at least~$\theta$.  In other
words, we can either fix the region confidence~$\theta$ and find the
largest region~$R_\diamond$ with confidence of at least~$\theta$, or we
can fix the minimum region size (support) $1000\eta$ and find the most
confident region~$R_*$ with support of at least $1000\eta$.

Observe that~$\tau$ in the definition of an optimized-gain admissible
region is the relative importance of support vs. that of confidence.
We can find a small region with high confidence or a large region with
small confidence, but both objectives cannot be maximized
simultaneously.  Increasing~$\tau$ will increase the confidence of the
optimized-gain admissible region, but decrease its support.  Likewise,
decreasing~$\tau$ will decrease the confidence of the optimized-gain
admissible region, but increase its support.  Therefore, we can find
approximate optimized-confidence and optimized-support admissible
regions by a binary search for the value of~$\tau$ where the respective
threshold is barely satisfied.  The search can stop at a given level of
precision $\Delta\tau$, where the lower bound on $\Delta\tau$ can be
found in~\cite{fukuda-rectilinear} (they show that the number of steps
in this search is logarithmic in the support $1000M$ of the bucket
space).  This algorithm is approximate because an optimized-confidence
(resp. optimized-support) admissible region need not be an
optimized-gain admissible region for any value of~$\tau$.  Yoda et
al.~\cite{fukuda-rectilinear} argue that this approximation is
reasonable for large datasets.

Let us revisit the definition of region `goodness'.  Geometrically, the
buckets with the same value of~$X$ are the columns and the buckets with
the same value of~$Y$ are the rows.  An optimized-gain admissible
region can be computed in $O(M_XM_Y^2)$ time by a set of rules of the
following form.  Recall that a region of type W gets wider from left to
right (see Figure~\ref{fig:admissible}).  Let $R_W(m,[s,t])$ be the
region of type W with maximum gain $f_W(m,[s,t])$ over all admissible
regions of type W that end in column~$m$ and span rows~$s$ through~$t$
in this column.  Then, either (a)~$m$ is the first column of
$R_W(m,[s,t])$, or (b)~$R_W(m,[s,t])$ includes the region
$R_W(m-1,[s',t'])$ with the maximum gain $f_W(m,[s',t'])$ over all
admissible regions that end in column~$m-1$ and span rows $s'\ge{}s$
through $t'\le{}t$ in this column.  \cite{fukuda-rectilinear}~keeps the
regions with maximum gain for every region type and every triple
$(m,[s,t])$ in a dynamic programming table.  These locally maximal
regions then grow according to a set of rules that compute an
optimized-gain admissible region.  This efficient greedy algorithm for
computing optimized-gain admissible regions depends on the property of
the gain function that we refer to as monotonicity.  Let
$0\le\tau\le{}1$ be a slope and~$R_{m'}$ and~$R_{m''}$ be two
admissible regions with gains $G(R_{m'},\tau)\ge{}G(R_{m''},\tau)$.
The gain function $G(R_m,\tau)$ is \emph{monotonic} if for any region
$R_k$ disjoint with both~$R_{m'}$ and~$R_{m''}$
$$G(R_{m'}\cup{}R_k,\tau)\ge{}G(R_{m''}\cup{}R_k,\tau),$$ where the
union of regions is defined in the obvious way.  It is easy to see that
our gain function
$$G(R_m,\tau)=H_m-\tau{}S_m=\sum_{i=1}^{\eta(m)}\lfloor{}1000P(E[B_{\kappa(m,i)}]<T)+0.5\rfloor-1000\tau\eta(m)$$
is monotonic because it is additive.  However, a strongly model-based
gain function $$G^{(M)}(R_m,\tau)=P(E[Q_m]<T)-\tau\eta(m)/M$$ is not
monotonic even if we assume independence of bucket random variables
$\{B_{\kappa(m,i)}\}_{i=1}^{\eta(m)}$ that make up~$Q_m$.  To the best
of our knowledge, only monotonic gain functions are known to result in
practical algorithms for computing optimized-gain admissible regions.

What happens when no estimates of mean and/or variance are available
for some bucket~$C_k$?  The answer to this question depends on
problem-specific considerations.  As was demonstrated in
Section~\ref{sec:w-example}, it is sometimes possible to provide
conservative estimates for these values.  For example, we have
empirically shown that the expected BEPs of some configurations
$\{c_k\}$ are smaller than $T=10^{-3}$ with confidence
$P(E[b_k]<T)\ge{}0.995$.  Likewise, we know that as the effective SNR
approaches negative infinity (in~dB), the BEP approaches 0.5, which is
the probability of correctly guessing the value of a random bit when
the transmitter is turned off.  Thus, we can let $P(E[b_k]<T)=0$ for
points with sufficiently small effective SNRs and a reasonable
performance threshold~$T$.  If no such estimates are available, we can
simply omit the missing buckets from the probability computation.  This
must be done with care because such buckets will contribute nothing to
the confidence of the region.  This fact can be used to reduce the
computational expense of sampling.

This section has highlighted the sometimes contradictory objectives
that aggregation must satisfy: permit valid statistical interpretations
and afford structure that can be exploited by data mining algorithms.
Our approach has been a judicious mix of concepts from both statistics
and data mining.  We showed that our formulation of the data mining
problem lies between the completely model-free approach and the
strongly model-based approach.  
The next section
applies the data mining methodology described here to the example in
Section~\ref{sec:w-example}.

\section{Optimized-Support Regions for the STTD Example}
\label{sec:experiments}

This section continues the example in Section~\ref{sec:w-example}.
First, we show that optimized-gain regions are both rectilinear and
connected for this example.  It immediately follows that
optimized-support and optimized-confidence regions are also
admissible.  An optimized-support admissible region is presented next.
We show that the elaborate region mining setup leads to simple
engineering interpretations.  Finally, we look at the performance of
data mining when the number of samples is small.  Three-fold
cross-validation shows that data mining performs well under these
circumstances.

\subsection{Justification of Data Mining for the STTD Example}

Let the average SNRs $S_1=X$ and~$S_2=Y$ partition the space of
configurations in Figure~\ref{fig:space} into disjoint points (buckets)
$\{c_k\}_{k=1}^M$, $1\le{}M\le{}1600$.
We now give an intuitive
argument to justify the suitability of the data mining algorithm for the
STTD study.
Without loss of generality,
consider only the points with $X\le{}Y$, i.e., $S_1\le{}S_2$.  It is
easy to extend all arguments to $X>Y$, but this adds little to the
discussion.

\begin{figure}
\begin{center}
\includegraphics[width=4.0in]{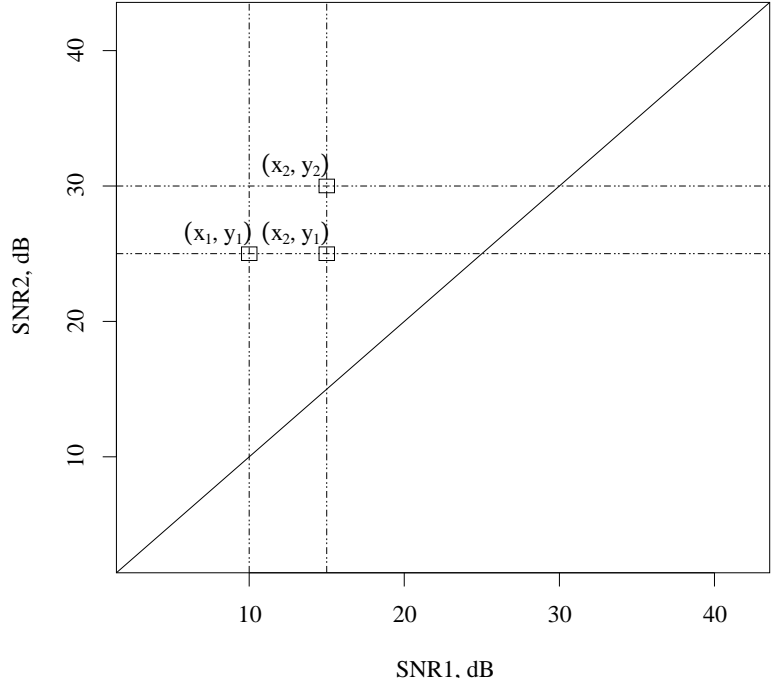}
\end{center}
\caption[Why optimal regions are admissible.]{Points for arguments
about region shape (see text).}
\label{fig:admissible-example}
\end{figure}

Let $c_1$ at $(x_1,y_1)$ and $c_2$ at $(x_2,y_1)$, $x_1<x_2<y_1$, be
two points in an optimized-gain region (of arbitrary shape) for some
slope $0<\tau<1$ (see Figure~\ref{fig:admissible-example}).  This means
that the confidences of these points are one, and thus the expected
BEPs of these points are smaller than the performance threshold~$T$.
When $x_1,x_2<y_1$ and $y_1$ is fixed, the BEP is a monotonically
decreasing function of $x$---increasing~$x$ decreases the power
imbalance and increases the effective SNR, so the BEP must decrease.
Therefore, the expected BEP of any point~$c_u$ at $(x_u,y_1)$,
$x_1<x_u<x_2$, is below the performance threshold~$T$.  Thus, the
confidences of points~$\{c_u\}$ are one and these points must also be in the
optimized-gain region~$R_\tau$.  Three more symmetric arguments of this
kind show that optimized-gain regions are rectilinear.

Likewise, let $c_1$ at $(x_1,y_1)$ and $c_2$ at $(x_2,y_2)$,
$x_1<x_2<y_1<y_2$, be two points in an optimized-gain rectilinear
region (refer to Figure~\ref{fig:admissible-example}).  Since $c_1$ is
in the optimized-gain region and $x_1<x_2$, the point at $(x_2,y_1)$ is
also in this region because it has a smaller BEP than~$c_1$.  Since the
optimized-gain region is rectilinear, there is a horizontal path from
$(x_1,y_1)$ to $(x_2,y_1)$ and a vertical path from $(x_2,y_1)$ to
$(x_2,y_2)$.  Thus, there is a Manhattan path from $(x_1,y_1)$ to
$(x_2,y_2)$.  Arguments of this kind show that optimized-gain
rectilinear regions must be connected as long as they are `wide
enough'.

To summarize, we have shown that 
optimized-gain (and thus optimized-support and optimized-confidence)
regions are admissible. The data mining
algorithm described in Section~\ref{sec:gizmo}, which results in
optimal admissible regions, is thus appropriate for the
STTD example.  We now show and interpret data mining results.

\subsection{Optimized-Support Admissible Regions}

\begin{figure}
\begin{center}
\includegraphics[width=4.0in]{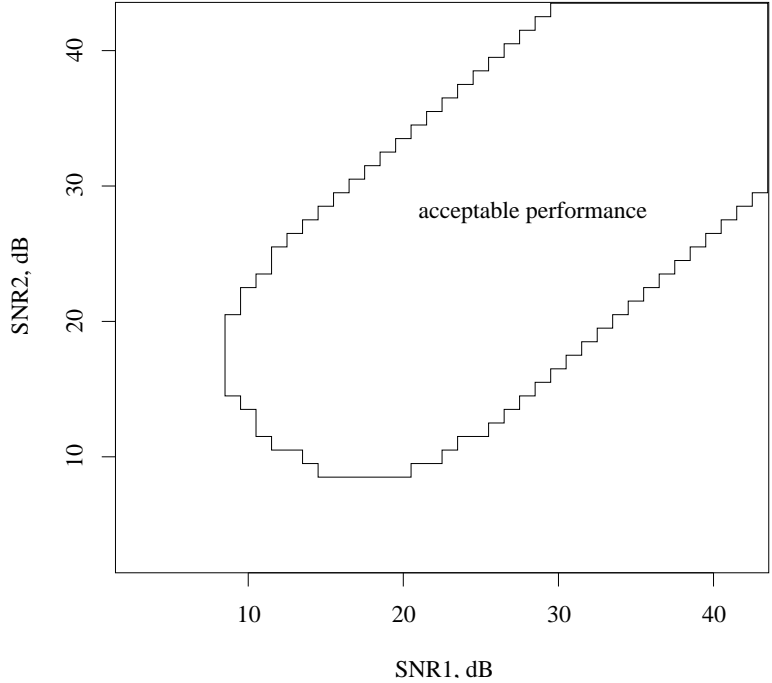} \\
\end{center}
\caption[Optimized-support admissible regions for
Figure~\ref{fig:space}.]{Optimized-support admissible region for data
in Figure~\ref{fig:space} (bottom) with the confidence
threshold~$\theta=0.99$ and the performance threshold~$T=10^{-3}$.}
\label{fig:region}
\end{figure}

Figure~\ref{fig:region} shows an optimized-support admissible region
for the confidence threshold~$\theta=0.99$.  Intuitively, this is the
largest admissible region where we can claim, with confidence of at
least~$0.99$, that configurations exhibit acceptable performance.  This
claim is conditional on temporal simulation assumptions and on mutual
independence of configurations in the region.  The shape of this region
confirms that, under a fixed effective SNR, the BEP is minimal when the
average SNRs of the two branches are equal.  The width of this region
shows the largest acceptable power imbalance.  For this example, the
system tolerates power imbalance of up to 12~dB.  However, the width of
the optimized region is not uniform.  The region is narrower for small
effective SNRs and wider for large effective SNRs.  This means that
configurations with low effective SNRs are more sensitive to power
imbalance than configurations with high effective SNRs.  None of these
observations are news to an informed reader.  The contribution of data
mining in this context is not qualitative discoveries; it is
statistically significant quantitative results.

\begin{figure*}
\begin{center}
\begin{tabular}{c c}
\includegraphics[width=250pt]{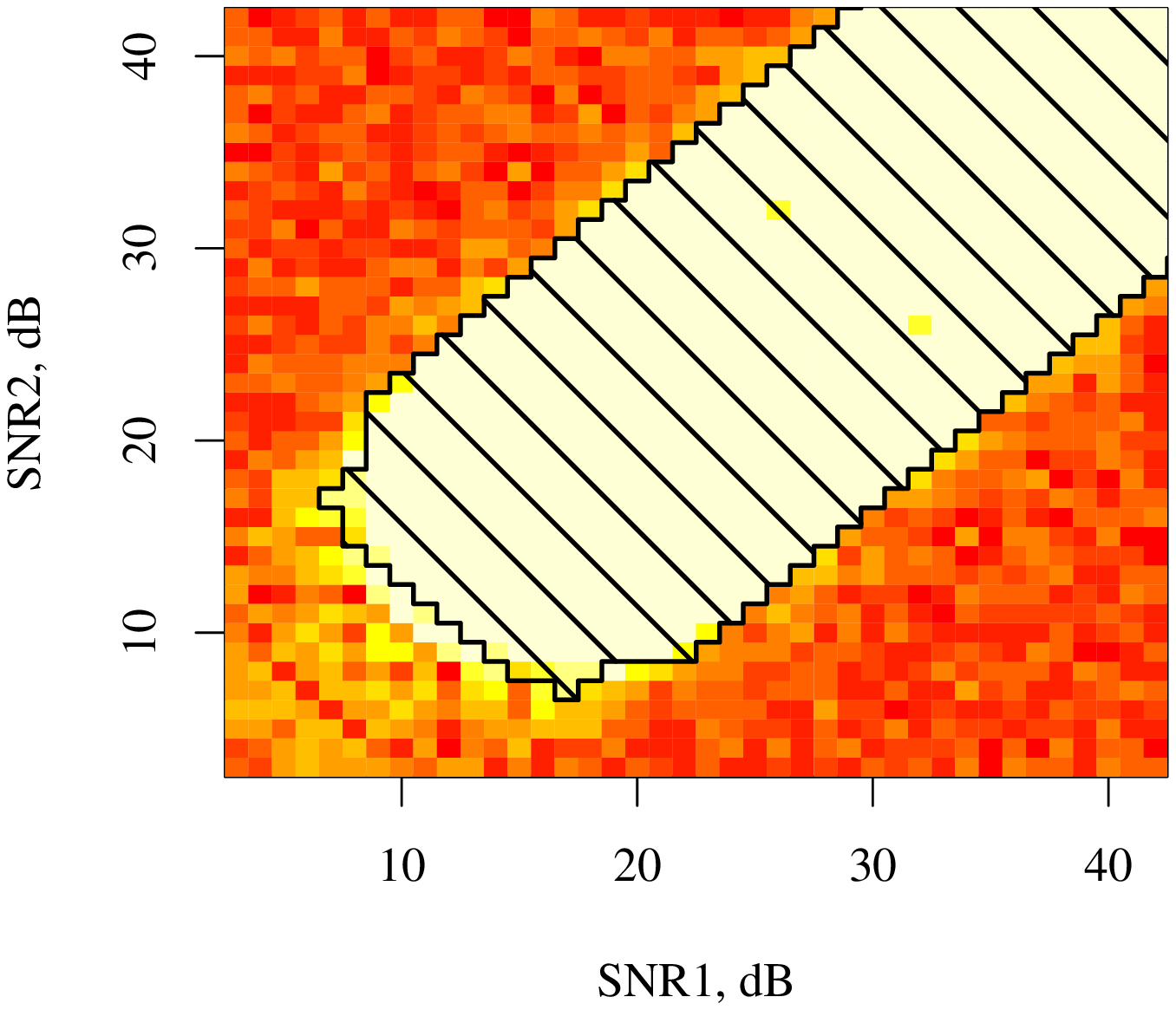} &
\includegraphics[width=250pt]{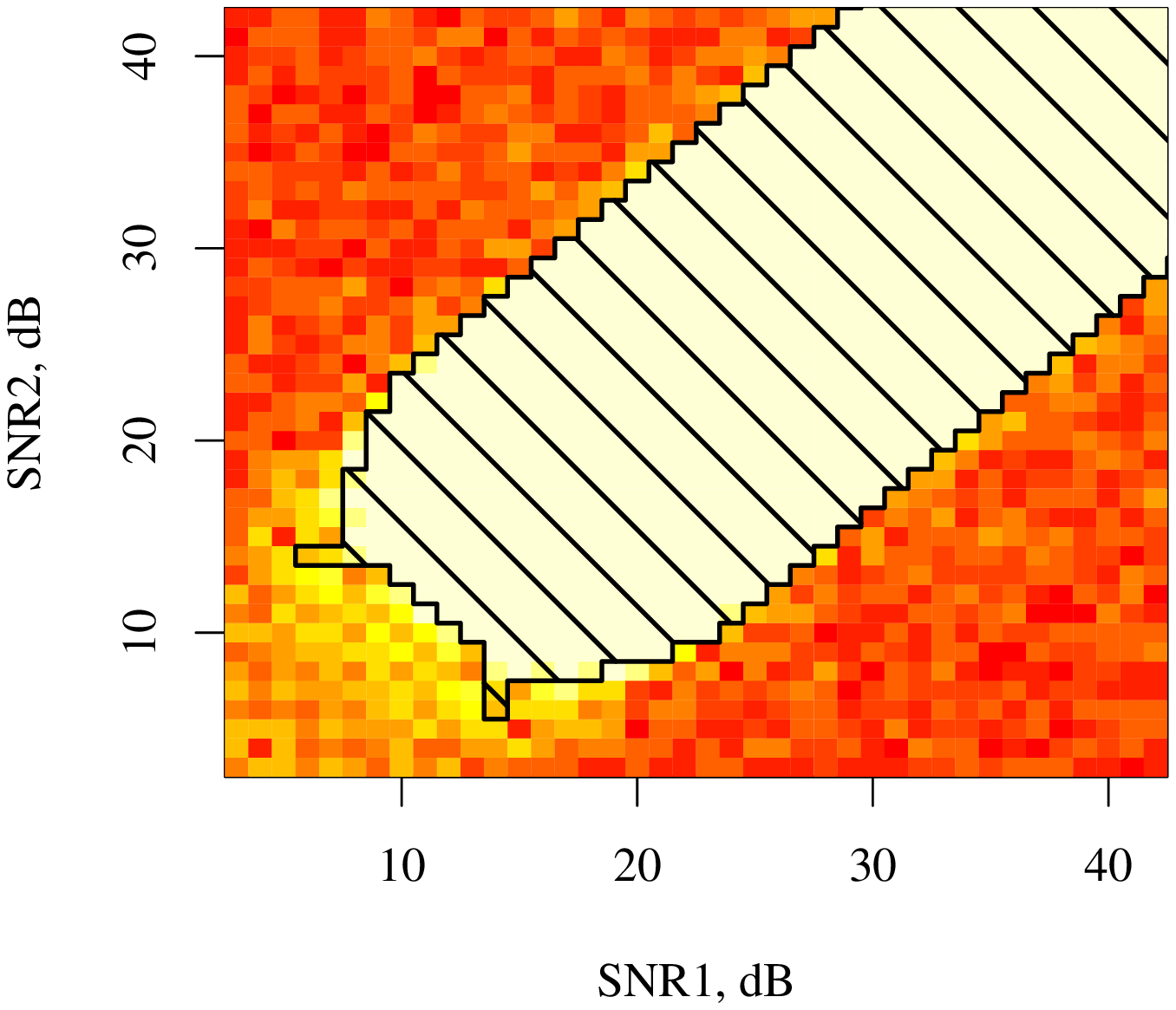} \\
\includegraphics[width=250pt]{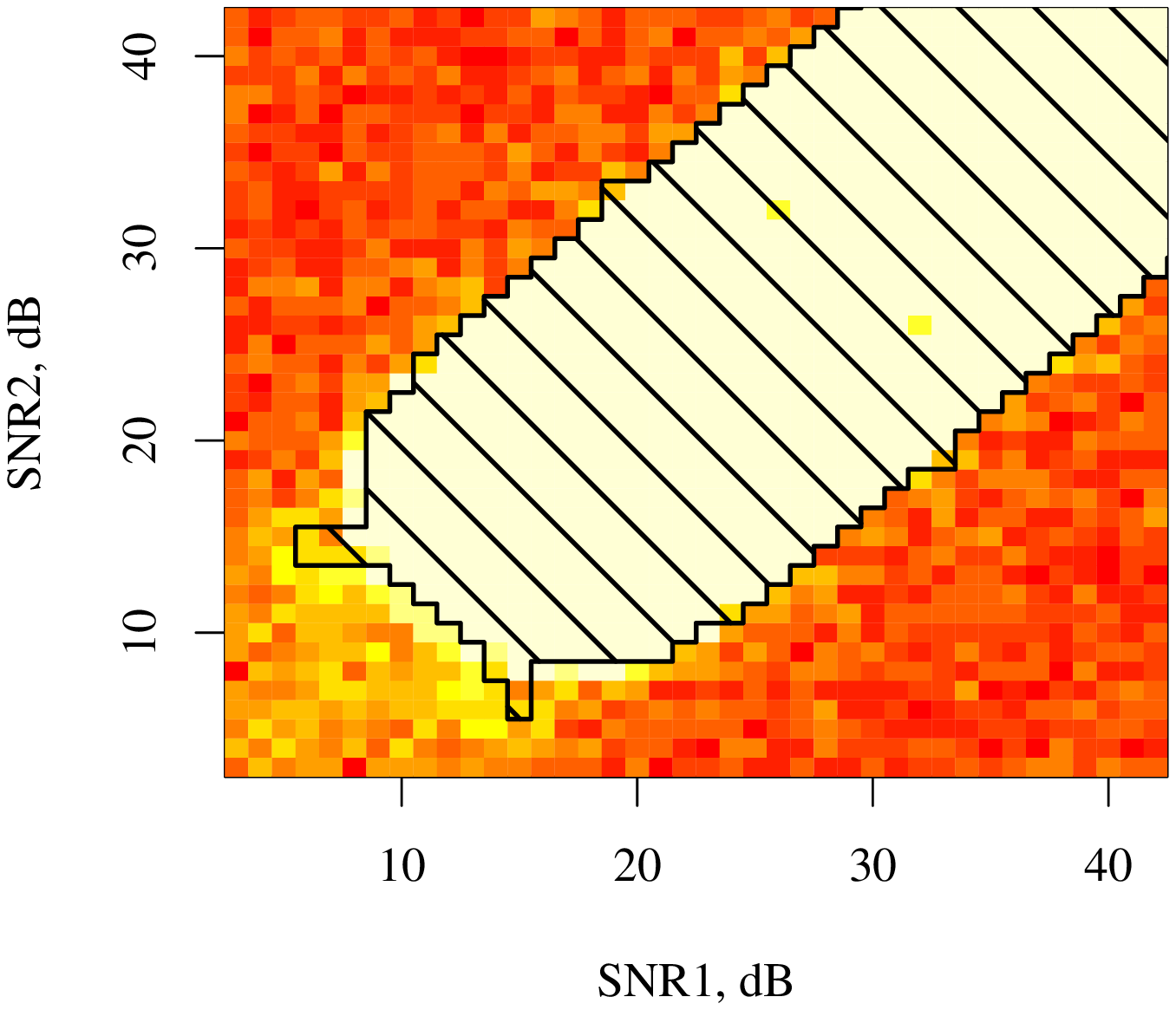} &
\includegraphics[width=250pt]{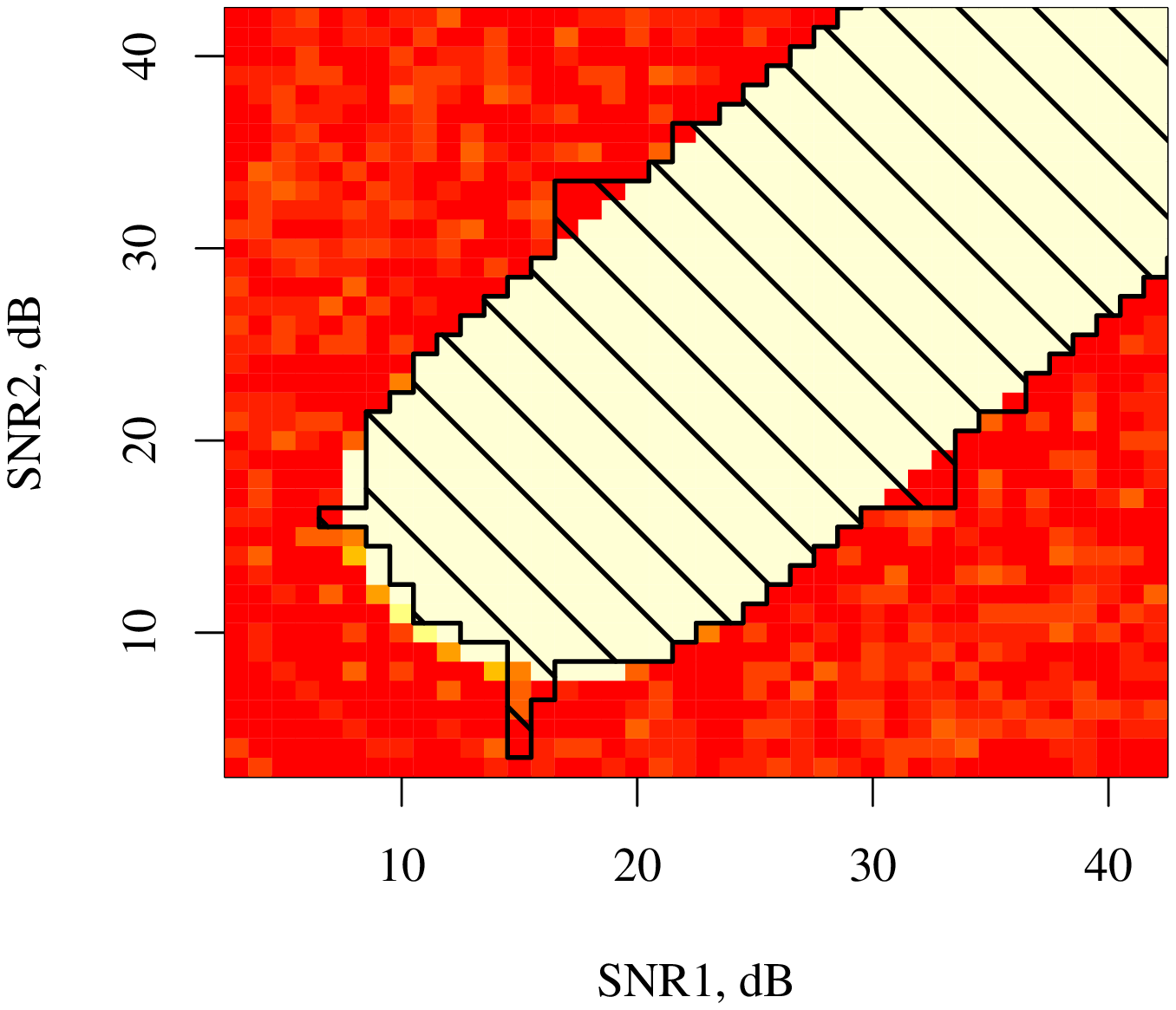} \\
\end{tabular}
\end{center}
\caption[Cross-validation of optimized-support admissible
regions.]{Cross-validation of optimized-support admissible regions with
the confidence threshold~$\theta=0.95$.  The regions in top left,
bottom left, and top right have been computed with $n_k=2$ independent
samples per bucket.  There are $758\pm{}2$ buckets ($47\%$ of all data)
per such region.  The region in the bottom right has been computed from
the statistically significant data in Figure~\ref{fig:space} (bottom).
It consists of 766 buckets ($48\%$ of all data).  Red (dark)
corresponds to low bucket confidence and white (light) corresponds to
high bucket confidence w.r.t. the voice quality threshold~$T=10^{-3}$.}
\label{fig:cv}
\end{figure*}

Let us see how the data mining algorithm performs when data is scarce.
The initial sample of the configuration space in Figure~\ref{fig:space}
(top) contains one sample value per bucket.  The statistically
significant sample in Figure~\ref{fig:space} (bottom) contains at least
two additional sample values per bucket (recall that we required at
least two sample values to estimate bucket variance $\sigma_k^2$).
Therefore, three-fold cross-validation is the most elaborate
cross-validation procedure that this dataset affords.  The regions in
the top left, bottom left, and top right of Figure~\ref{fig:cv} have
been computed for sample values in Figure~\ref{fig:space} (top) and the
first two sample values per bucket in Figure~\ref{fig:space} (bottom).
Each of these regions has been computed with two out of the three
sample values per bucket.  The region in the lower right has been
computed with all data in Figure~\ref{fig:space} (bottom).  All four
regions are optimized-support admissible regions with the confidence
threshold~$\theta=0.95$.  The regions are overlaid on top of the
color-coded bucket confidence values.  Red (dark) corresponds to low
confidence and white (light) corresponds to high confidence that
configuration $c_k$ exhibits acceptable average performance w.r.t. the
voice quality threshold~$T=10^{-3}$.

The regions in Figure~\ref{fig:cv} are identical except for the lower
left corner.  This is not surprising because this part of the
configuration space exhibits high relative variance.  Also, the data is
symmetric but the regions are asymmetric in the lower-left corner.
Recall that optimized-gain admissible regions, and thus
optimized-support admissible regions, are not unique.  The ties in
region gains are broken arbitrarily.  Therefore, region asymmetry is an
additional indicator of region instability.

Figure~\ref{fig:cv} also shows that additional data improves image
contrast but does not significantly affect region shape.  Collecting
additional sample values separates the points into ones with low
confidence and ones with high confidence.  A curious side effect occurs
when the difference in confidence estimates of low-confidence points
falls below the discretization error (1/1000).  In this case, the
`confidence slack' $1-\theta$ is allocated to arbitrary points with low
confidence.  One way to correct this situation is to raise the
confidence threshold~$\theta$---after all, more accurate data should
afford stronger claims.  Another alternative is to lower the
discretization threshold.  In general, optimized regions work best when
the data is noisy.  A contour plot will suffice when the data is highly
accurate.

It can also be seen that the high contrast created by the sharp edge of
the tolerance region is advantageous to data mining.  The region is
stable where the contrast is high.  When the image is blurred, data
mining tries to avoid the questionable boundary points.

To summarize, this section has demonstrated that optimized-gain regions
are rectilinear and connected for a non-trivial space of wireless
system configurations.  We have also shown that optimized-support
admissible regions are easy to interpret.  Finally, we have shown that
data mining works well when sample sizes are small.

\section{Discussion and Future Work}
\label{sec:conclusion}

We have demonstrated a hierarchical formulation of data mining suitable
for assessing performance of wireless system configurations.  WCDMA
simulation results are systematically aggregated and redescribed,
leading to intuitive regions that allow the engineer to evaluate
wireless system configuration parameters.  We have shown that the
assumptions about region shape and properties made by data mining
algorithms can be valid in the wireless design context; the patterns
mined hence lead to explainable and statistically valid design
conclusions.  As a methodology, data mining is thus shown to be
extremely powerful when coupled with statistically meaningful
performance evaluation.

This work is the first (known to the authors) application of data
mining methodology to solve problems in wireless system design.
Therefore, a large number of extensions are possible and called for.
We outline possible extensions at the three levels of aggregation:
points, buckets, and regions.

At the point level, it may be advantageous to model temporal
simulations more precisely.  This paper assumes a `large enough' number
of frames per simulation and works with the distribution of estimated BEPs.
 We have shown reasonable analytical and empirical evidence that this
distribution is Gaussian.  The advantage of this problem formulation is
the independence of spatial aggregation from the assumptions of
temporal simulation.  This helps introduce wireless engineers to the
methodology of data mining for studying design problems.  However, a
stronger model of temporal simulation (e.g., Markov chains
in~\cite{fsmc-wang}) may yield appreciable gains in software
performance.  This direction is worth pursuing because few research
groups have access to parallel computing facilities of the scale used
in this work.  For instance, the initial sample of the configuration
space in Figure~\ref{fig:space} (top) would take one year of
computation time on a modern workstation.  The study presented in this
paper would clearly be impossible without significant computational
power.

Aggregation of points into buckets is the least developed part of this
work.  Suppose that we would like to simulate the effects of
interference on configuration performance.  Assume that the
distribution of the average strengths of the interfering signals is
known \emph{a priori} (e.g., estimated by ray tracing).  We can either
make this distribution known to the temporal simulation, or,
alternatively, run several temporal simulations for different strengths
of interfering signals.  The former is more accurate and
computationally more efficient, but the latter is more generic and
simpler to implement.  Bucketing of simulation results with varying
simulation parameters is intended to approximate the performance of a
single device under varying conditions.  This paper does not employ
such bucketing but instead builds all the necessary kinds of parameter
variation into the temporal simulation (which can be argued to be the
right way to do it).  However, bucketing may be necessary when one has
to work with a given dataset (e.g., measurements).  Bucket space can be
viewed as a configuration space for a more complex temporal
simulation.  Therefore, an in-depth treatment of bucketing is
orthogonal to the primary topic of this paper, which is data mining.

Significant work remains to be done at the region level as well.  For
instance, the assumption of small variance could conceivably be relaxed.
One can
also pursue the relatively difficult task of incorporating strongly
model-based prior knowledge into the data mining algorithm, or the
somewhat easier task of applying different kinds of region mining
algorithms to problems in wireless system design.

Defining additional case studies is another obvious direction for
future work.  We have studied a relatively small part of the parameter
space of modern wireless systems.  More studies of this type must be
performed to highlight the merits and the shortcomings of data mining
in this domain.

Finally, the strict staging of data collection and data mining can be
relaxed.  One can fruitfully interleave the two activities and have the
results of data mining drive subsequent data collection.  In
data-scarce domains, it would be advantageous to focus the data
collection effort on only those regions deemed most important to
support a particular data mining objective.  Methodologies for
closing-the-loop in this manner are becoming increasingly
prevalent~\cite{sampling-cise}. This will also help define alternative
criteria for evaluating experiment designs and layouts.

\bibliographystyle{alpha}
\bibliography{paper}

\newcommand{\etalchar}[1]{$^{#1}$}
\ifx\etalchar\undefined\else\renewcommand{\etalchar}[1]{}\fi
\begin{thebibliography}{VHW{\etalchar{+}}02}

\bibitem[Ala98]{sttd-alamouti}
S.M. Alamouti.
\newblock {A Simple Transmit Diversity Technique for Wireless Communications}.
\newblock {\em IEEE Journal on Selected Areas in Communications}, Vol.
  16(8):pages 1451--1458, October 1998.

\bibitem[CB02]{stat-casella}
G.~Casella and R.L. Berger.
\newblock {\em Statistical Inference}.
\newblock Duxbury, 2nd edition, 2002.

\bibitem[DDJS02]{sttd-stutzman}
K.~Dietze, C.B. Dietrich~Jr., and W.L. Stutzman.
\newblock {Analysis of a Two-Branch Maximal Ratio and Selection Diveristy
  System With Unequal {SNRs} and Correlated Inputs for a {R}ayleigh Fading
  Channel}.
\newblock {\em IEEE Transactions on Wireless Communications}, Vol. 1(2):pages
  274--281, April 2002.

\bibitem[FMMT01]{fukuda-tods}
T.~Fukuda, Y.~Morimoto, S.~Morishita, and T.~Tokuyama.
\newblock {Data Mining with Optimized Two-Dimensional Association Rules}.
\newblock {\em ACM Transactions on Database Systems}, Vol. 26(2):pages
  179--213, June 2001.

\bibitem[FPSS96]{kdd-cacm}
U.M. Fayyad, G.~Piatetsky-Shapiro, and P.~Smyth.
\newblock {The {KDD} Process for Extracting Useful Knowledge from Volumes of
  Data}.
\newblock {\em Communications of the ACM}, Vol. 39(11):pages 27--34, November
  1996.

\bibitem[Has93]{cm-hashemi}
H.~Hashemi.
\newblock {The Indoor Radio Propagation Channel}.
\newblock {\em Proceedings of the IEEE}, Vol. 81(7):pages 943--968, July 1993.

\bibitem[HT00]{wcdma-holma}
H.~Holma and A.~Toskala.
\newblock {\em WCDMA for UMTS: Radio Access for Third Generation Mobile
  Communications}.
\newblock John Wiley, New York, 2000.

\bibitem[HTF01]{dm-stat}
T.~Hastie, R.~Tibshirani, and J.H. Friedman.
\newblock {\em The Elements of Statistical Learning: Data Mining, Inference,
  and Prediction}.
\newblock Springer Verlag, October 2001.

\bibitem[JBS92]{comm-jeruchim}
M.C. Jeruchim, P.~Balaban, and K.S. Shanmugan.
\newblock {\em Simulation of Communication Systems}.
\newblock Plenum Press, New York, 1992.

\bibitem[RBK02]{sampling-cise}
N.~Ramakrishnan and C.~Bailey-Kellogg.
\newblock {Sampling Strategies for Mining in Data-Scarce Domains}.
\newblock {\em IEEE/AIP Computing in Science and Engineering}, Vol. 4(4):pages
  31--43, July/August 2002.

\bibitem[RG01]{naren-ayg}
N.~Ramakrishnan and A.Y. Grama.
\newblock {Mining Scientific Data}.
\newblock {\em Advances in Computers}, Vol. 55:pages 119--169, September 2001.

\bibitem[UMT97]{umts-utra}
{Universal Mobile Telecommunications Systems (UMTS) UMTS Terrestrial Radio
  Access System (UTRA) Concept Evaluation}.
\newblock Technical Report 101 146 v3.0.0, ETSI, Sophia Antipolis, France,
  December 1997.

\bibitem[VHW{\etalchar{+}}02]{ipdps-s4w}
A.~Verstak, J.~He, L.T. Watson, N.~Ramakrishnan, C.A. Shaffer, T.S. Rappaport,
  K.~Anderson, C.R.~Bae, J.~Jiang, and W.H. Tranter.
\newblock {{S$^4$W}: Globally Optimized Design of Wireless Communication
  Systems}.
\newblock In {\em Proceedings of the Next Generation Software Workshop, 16th
  International Parallel and Distributed Processing Symposium (IPDPS'02)}. IEEE
  Press, April 2002.
\newblock Fort Lauderdale, FL.

\bibitem[Wan95]{fsmc-wang}
H.S. Wang.
\newblock {Finite-State {M}arkov Channel---A Useful Model for Radio
  Communication Channels}.
\newblock {\em IEEE Transactions on Vehicular Technology}, Vol. 44(1):pages
  163--171, February 1995.

\bibitem[YFM{\etalchar{+}}97]{fukuda-rectilinear}
K.~Yoda, T.~Fukuda, Y.~Morimoto, S.~Morishita, and T.~Tokuyama.
\newblock {Computing Optimized Rectilinear Regions for Association Rules}.
\newblock In {\em Proceedings of the Third International Conference on
  Knowledge Discovery and Data Mining (KDD'97)}, pages 96--103, August 1997.

\bibitem[ZT02]{comm-ziemer}
R.E. Ziemer and W.H. Tranter.
\newblock {\em Principles of Communications: Systems, Modulation, and Noise}.
\newblock John Wiley, New York, 5th edition, 2002.

\end{thebibliography}
\end{document}